\documentstyle[prl,aps]{revtex}
\input{epsf.sty}

\def\Tr{{\rm Tr}}
\def\L{{\cal L}}

\def\r{{\bf r}}

\def\bp{{\bf p}}
\def\bk{{\bf k}}
\def\bD{{\bf D}}
\def\b\mu{{\bf \mu}}

\def\cD{{\cal D}}

\begin{document}

\draft

\title{The Chiral Fermion Meson Model at Finite Temperature}
\author{H. C. G. Caldas$^{\dagger}$$^{\ddag}$\thanks{e-mail: hcaldas@funrei.br}, A. L. Mota$^{\dagger}$
and M. C. Nemes$^{\ddag}$}

\address{${\dagger}$Departamento de Ci\^{e}ncias Naturais, DCNAT \\
Funda\c{c}\~ao de Ensino Superior de S\~{a}o Jo\~{a}o del Rei, FUNREI,\\ 
Pra\c{c}a Dom Helv\'ecio, 74, CEP:36300-000, S\~{a}o Jo\~{a}o del Rei, MG, Brazil} 

\address{${\ddag}$Departamento de F\'\i sica, Universidade Federal de Minas Gerais,\\
CP 702,CEP:30.161-970,Belo Horizonte, MG, Brazil}
\date{September, 2000}

\maketitle                          

\begin{abstract}
We study the chiral fermion meson model which is the well known linear 
sigma model of Gell-Mann-and-Levy  at finite temperature.
A modified self-consistent resummation (MSCR) which resums higher order 
terms in the perturbative expansion is proposed. It is shown that with 
the MSCR the problem of tachyonic masses is solved, the renormalization 
of the gap equations is carried out and the Goldstone's theorem is 
verified. We also apply the method to investigate another known case 
at high temperature and compare with results found in the literature.

\end{abstract}


\section{Introduction} 

Several models have been proposed to describe hadron properties in the 
regime of low energies. Among these models, we adopt the linear $\sigma$ 
model of Gell-Mann and Levy\cite{Gell-Mann} which is a phenomenological 
model of Quantum Chromodynamics-QCD that incorporates two important 
features of QCD: chiral symmetry and partial conservation of axial vector 
current. The model was originally proposed as a model for strong interactions
\cite{Gell-Mann}, but nowadays it serves as an effective model for the low 
energy (low temperature) phase of QCD. It has the advantage of being renormalizable 
at zero\cite{Franco} and finite temperature\cite{Ram-Mohan}. Altough the linear 
sigma model lagrangian  exhibits chiral symmetry, quantum effects break this symmetry 
spontaneously. Both from theoretical\cite{Lattice} and  experimental \cite{Quark} point 
of view, there exist a great amount of interest in the study of  chiral 
symmetry restoration at finite temperature.

However, quantum field theory at high temperature has a well known problem 
that is the breakdown of the perturbative 
expansion\cite{Dolan,Weinberg,Linde}. This happens in theories with 
spontaneous symmetry breaking (SSB) or in massless field theories because 
powers of the temperature can compensate for powers of the coupling constant. Resummation techniques which try consistently to take into account 
higher-loops are required.

A systematic self-consistent approximation approaches  based on the meson 
sector of the linear $\sigma$ model was previously studied by Baym and Grinstein\cite{Baym}. After, Banerjee and Mallik\cite{Mallik} proposed a modified perturbation expansion with the objective of calculating the two-point functions up to second order in the $\lambda \phi^{4}$ theory. A resummed perturbative expansion was proposed by Parwani \cite{Parwani} in order to go beyond leading order in the same model. More recently, Chiku and Hatsuda\cite{Chiku} in the study of the $O(N)$ $\phi^{4}$ model presented a novel resummation adding a mass parameter determined later by the fastest apparent convergence (FAC) condition. We employ imaginary-time formulation (ITF) whereas in \cite{Chiku} real-time formulation (RTF) is used in the development 
of the optimized perturbation theory (OPT). In this paper we develop a 
modified self-consistent resummation at finite temperature and apply it 
to the investigation of the chiral fermion meson model. We study the 
temperature dependence of the chiral condensate and the effective meson 
and fermion masses by this self-consistent non-perturbative approximation 
up to one-loop order in the perturbative expansion. In the application of 
the MSCR to the study of the chiral fermion meson model at finite 
temperature, we divided the problem into three physical regions: low, intermediate and high temperatures. This is essential to identify the 
regions where resummation is crucial. In each region renormalization and satisfaction of Goldstone's theorem are discussed in detail. Our study addresses 
problems found in the context of the well studied $O(4)$ linear sigma model 
and deals with a usually avoided point: the inclusion of the fermions.
Also, we re-examine the chiral phase transition in static equilibrium in 
terms of the linear sigma model with our MSCR. Instead of demanding a 
infinite gap-equation, as has been done often in the recent literature, 
we perform the renormalization in stages in order to get finite 
gap-equations.

We also treat an explicit chiral symmetry breaking term in the Lagrangian 
which generates the realistic finite pion mass. Symmetry is never restored in 
this case. It is shown that in the limit of vanishing pion mass, namely when 
the chiral symmetry is exact, the inclusion of fermions does not change the order (nature) of the phase transition but only lowers the value of the 
critical temperature. 

This paper is organized as follows. In Section II we discuss the chiral 
fermion meson model and some of its features at zero temperature. 
In Section III the temperature is introduced via the partition function 
of the model which lead to the thermodynamical potential. The inclusion 
of loop corrections and the thermal gap equations are addressed to 
Section IV. In section V we apply the MSCR to the study of the massless 
$\lambda \phi^4$ model in the weak coupling limit at high temperature. 
The renormalization of the self-energy is studied in 
Section VI. The numerical results are presented in Section VII. 
Section VIII is devoted to conclusions.

\section{The Chiral Fermion Meson Model at Zero Temperature}

The Lagrangian density of the chiral fermion meson model which provides an explicit realization of chiral symmetry is given by\cite{Gell-Mann}
\begin{equation}
\label{lagsy}
{\cal L}_{sym}  =  \overline{\psi}\left[i \gamma^{\mu} \partial_{\mu} - g(\sigma' +
 i \gamma^5{\vec\pi}\cdot {\vec\tau})\right]\psi +\frac{1}{2}\left[(\partial
 \sigma')^2 + (\partial \vec \pi)^2\right] \nonumber \\
 - \frac{\lambda}{4}(\sigma'^2 + \vec\pi^2 - f_\pi^2 )^2 ,\\
\end{equation}
where $\psi$, $\sigma'$, and $\pi$ represent the quark, sigma and
pion fields, respectively, $\lambda$ and $g$ are positive coupling constants 
and $f_\pi$ is the pion decay constant in vacuum.

If the up and down quark masses were zero, QCD would have a chiral $SU(2)_L \times SU(2)_R$  symmetry. In the vacuum this symmetry is spontaneously broken by quantum effects, with the result that there exists a triplet of Goldstone bosons. In reality the quark masses are very small but nonzero, so that chiral symmetry is only approximate and the pion has a small mass \cite{Kapusta}. An explicit chiral symmetry breaking term is added to the Lagrangian which generates the realistic finite pion mass so that

\begin{equation}
{\cal L' } = {\cal L}_{sym} + {\cal L}_{sym b}
\label{lagd}
\end{equation} 
with
\begin{equation}
{\cal L}_{sym b}  =   c \sigma'\\,
\label{laga}
\end{equation}
where $c$ is small and positive.

The term ${\cal L}_{sym}$ is symmetric and
invariant under an $SU(2)_L \times SU(2)_R$ chiral group and ${\cal
L}_{sym b}$ is the symmetry breaking term.  Two Noether currents
associated with (\ref{lagsy}), namely the vector current and the axial
vector current, are given by
\begin{eqnarray*}
\vec{V}_{\mu} =  \overline{\psi} \gamma_{\mu} \frac{\vec{\tau}}{2} \psi +
\vec{\pi} \times \partial_{\mu} \vec{\pi},  \\
\vec{A}_{\mu} =  \overline{\psi} \gamma_{\mu} \gamma_5
\frac{\vec{\tau}}{2} \psi + 
\sigma' \partial_{\mu} \vec{\pi} - \vec{\pi} \partial_{\mu} \sigma' 
\end{eqnarray*}  
respectively.  The equations of motion for the fields derived from the
Lagrangian density (\ref{lagsy}) give the PCAC relations
\begin{equation}
\partial_{\mu} \vec{A}^{\mu} = c \vec{\pi} .
\label{pcac}
\end{equation}

The effect of the term $c \sigma'$ on the classical fundamental state, can be found by looking at the minimum of the potential

\begin{equation}
V_{0}(\sigma',\vec{\pi})= \frac{\lambda}{4}(\sigma'^2 +
 \vec\pi^2 - f_\pi^2 )^2 - c \sigma'
\label{pot}
\end{equation}

\begin{equation}
\frac{\partial V_{0} (\sigma',\vec{\pi})} {\partial \sigma'}=
\lambda(\sigma'^2 + \vec\pi^2 - f_\pi^2) \sigma'-c = 0
\label{potd1}
\end{equation}

\begin{equation}
\frac{\partial V_{0} (\sigma',\vec{\pi})} {\partial \pi^a}=
\lambda(\sigma'^2 + \vec\pi^2 - f_\pi^2) \pi^a = 0
\label{potd2}
\end{equation}
whose (unique) solutions are

\begin{eqnarray}
\label{sol1}
\vec\pi_{0}=0  ,\\
\lambda(\sigma_{0}'^2 - f_\pi^2) \sigma_{0}' = c
\nonumber 
\end{eqnarray}

To first order in $c$, we have

\begin{equation}
\sigma_{0}' = f_\pi+ \frac{c}{2 \lambda f_\pi^2} \equiv \nu
\label{sol2}
\end{equation}

From (\ref{sol2}), we see that $\sigma_{0}'$ has a non-zero vacuum expectation value. It is convenient to redefine the sigma field as $\sigma' \to \sigma + \nu$ such that $\sigma$ has zero expectation value. As an effect of this shift the fermion field acquires a mass given by

\begin{equation}
m_{\psi}=g\nu
\label{massf}
\end{equation}

The shifted Lagrangian, ${\cal L}_{s}$, of the new quantum theory reads
\begin{eqnarray}
{\cal L}_{s}  = - \frac{\lambda}{4}(f_{\pi}^2- \nu^2)^2 + c\nu  - 
\lambda(\nu^3- \nu f_{\pi}^2- \frac{c}{\lambda}) \sigma 
\nonumber
+ \\ \overline{\psi}[i \gamma^{\mu} \partial_{\mu}  -m_{\psi}] \psi
 + \frac{1}{2} [(\partial \vec \pi)^2 -m_{\pi}^2 \vec \pi ^2 + (\partial \sigma)^2 -m_{\sigma}^2 \sigma ^2] 
\nonumber
+ \\ -g \overline{\psi}[ \sigma +i \gamma^5{\vec\pi}\cdot {\vec\tau})]\psi 
\nonumber 
- \frac{\lambda}{4}[(\vec\pi^2 + \sigma^2)^2 +4 \nu \sigma (\vec\pi^2 + \sigma^2)] = \\
- U(\nu) + {\cal L}_{0} + {\cal L}_{I}
\label{lagsh}
\end{eqnarray}
where $U(\nu)$ is the mean field energy density, ${\cal L}_{0}$ is the free Lagrangian and ${\cal L}_{I}$ is the interaction Lagrangian, defined by

\begin{equation}
U(\nu) \equiv \frac{\lambda}{4}(f_{\pi}^2-\nu^2)^2 -c\nu,
\label{enerdens}
\end{equation}

\begin{equation}
{\cal L}_{0} \equiv \overline{\psi}[i \gamma^{\mu} \partial_{\mu}  -m_{\psi}] \psi + \frac{1}{2} [(\partial \vec \pi)^2 -m_{\pi}^2 \vec \pi ^2 + (\partial \sigma)^2 -m_{\sigma}^2 \sigma ^2],
\label{lagfree}
\end{equation}

\begin{equation}
{\cal L}_{I} \equiv -g \overline{\psi}[ \sigma +i \gamma^5{\vec\pi}\cdot {\vec\tau})]\psi - \frac{\lambda}{4}[(\vec\pi^2 + \sigma^2)^2 +4 \nu \sigma (\vec\pi^2 + \sigma^2)],
\label{laginte}
\end{equation}
respectively.

The meson masses read out of the shifted Lagrangian (\ref{lagsh}) are

\begin{equation}
m_{\pi}^2=m^2+\lambda\nu^2,
\label{massp}
\end{equation}

\begin{equation}
m_{\sigma}^2=m^2+3\lambda\nu^2
\label{masss}
\end{equation}
where $m^2=-\lambda f_{\pi}^2 <0$.

It is easy to see that the coefficient of the linear term in the sigma field, $\lambda(\nu^3- \nu f_{\pi}^2- \frac{c}{\lambda}) $, in Lagrangian 
(\ref{lagsh}) is identically zero by the minimal condition (\ref{sol1}). 
This is due to the fact that the vacuum expectation value of the sigma field, $\langle \sigma \rangle $, should vanish at any order of perturbation theory\cite{Lee}, even if we include thermal corrections\cite{Kirzhnits}. 
The one-loop thermal tadpole corrections will modify this relation which 
will become temperature dependent.  If $\nu$ is allowed to be temperature dependent, the masses are temperature dependent as well. At any temperature, $\nu$ is such that $\langle \sigma \rangle =0$. At zero temperature, when $c$ continuously approaches zero, we have the solutions 
$\langle \vec\pi \rangle =0$ and $\langle \sigma' \rangle =f_{\pi}$ which minimize the potential satisfying the Goldstone's theorem.

The contact with phenomenology is made by fixing the parameters of the 
model to agree with the observable value of the particle masses in vacuum. 
Then, the tree level parameters of the Lagrangian are

\begin{equation}
\lambda=\frac{ m_{\sigma;0}^2-3m_{\pi;0}^2}{2 f_\pi^2},
\label{vlamb}
\end{equation}

\begin{equation}
c=\nu m_{\pi;0}^2 \simeq f_{\pi}m_{\pi;0}^2,
\label{vc}
\end{equation}

\begin{equation}
g=\frac{ m_{\psi;0} (m_{\sigma;0}^2-3m_{\pi;0}^2)}{ f_{\pi} ( m_{\sigma;0}^2-2m_{\pi;0}^2)},
\label{vg}
\end{equation}
where $m_{\pi;0}=139MeV$ , $m_{\sigma;0}=600MeV$ , $m_{\psi;0}=340MeV$  and $f_\pi=93MeV$.

As we have mentioned earlier, our goal in this work is to study the chiral phase transition in the chiral fermion meson model and to analyze the thermal behavior of the temperature dependent meson condensate $\nu$ and the meson and fermion masses. So it will be necessary to compute all the one-loop self-energies for the particles present in the model. Such self-energy diagrams have divergent pieces which must be renormalized if we want reliable results. In most of the approximations found in the literature\cite{Baym,Camelia1,Camelia2,Roh1} several difficulties have been found in the tentative of renormalizing the divergent gap-equations. Sometimes the undesirable  parts have been ignored\cite{Petropoulos}. The renormalization of the self-energy is studied in section\ref{Ren} whereas the effective potential renormalization is performed in appendix \ref{apend1}. For the purpose of renormalization it is necessary to add to ${\cal L}_{s}$  a counterterm Lagrangian, $L_{C.T.}$, needed to render the theory finite \cite{Pierre}, 
\begin{equation}
{\cal L}={\cal L}_{s} + L_{C.T.}
\label{lagr}
\end{equation}
where
\begin{equation}
L_{C.T.} = C.T. + D_{1} m_{\psi}^4 + D_{2} m_{\pi}^4 + D_{3} m_{\sigma}^4
\label{lagcot}
\end{equation}

In (\ref{lagcot}) $C.T.$ contains the appropriate counterterms to be used in the renormalization of the masses while $D_{1,2,3} m_{\psi,\pi,\sigma}^4$ are necessary to keep the thermodynamical potential finite, as we will see further. As we are interested only in the study of the thermal effective masses at one loop order in the perturbative expansion, other counterterms necessary to renormalize the coupling constants are not explicitly shown.

\section{The Partition Function of the Model and the link to Statistical Mechanics}

One of the most fundamental objects in thermodynamics is the partition function, defined by,

\begin{equation}
\label{partfu}
Z = \Tr \  e^{ - \beta \hat H },
\end{equation}
where $\hat H$ is the Hamiltonian of the system, $\beta = 1/ k_B
T=T^{-1}$ with the Boltzmann constant, $k_B$, set equal to one and the trace, $\Tr$, in eq. (\ref{partfu}) meaning the sum of the elements of the matrix $ e^{ - \beta \hat H }$ in all independent states of the system. All information concerning the equilibrium thermodynamic macroscopic properties of the system are obtained from $Z$.

In relativistic quantum field theory, the partition function can be derived from the Feynman's functional formalism \cite{Bernard}. The bridge between quantum mechanics and statistical mechanics is achieved by the heuristic introduction of a variable defined as $\tau=it$. Also, the fields are constrained to obey periodic(anti) boundary conditions: $\phi (\r, 0) = \phi(\r, \beta)$ for bosons and  $\psi (\r, 0) = - \psi(\r, \beta)$ for fermions. Following these prescriptions, we get

\begin{eqnarray}
\label{partfuc}
Z[\overline{\psi},\psi,\sigma,\pi] = N' \int  \cD [ \phi ]  \exp [ \int_0^{\beta} d\tau \int d^3 x  \L (\phi, \partial \phi) ] 
= \\ \exp \int_\beta d^4x \left[-\frac{\lambda}{4}(f_{\pi}^2-\nu^2)^2 +c\nu + D_{1} m_{\psi}^4 + D_{2} m_{\pi}^4 + D_{3} m_{\sigma}^4\right] N' \int  \cD [ \phi ] e^{S_0} e^{S_I}
\nonumber
\end{eqnarray}
Here we have introduced a short hand notation for the Euclidean space-time integral: $S= \int_\beta d^4x \L \equiv \int_0^{\beta} d\tau \int d^3 x \L$, $\cD [ \phi ]$ is an abbreviation for the integral over $\overline{\psi}$, $\psi$, $\sigma$ and $\pi$, $N'$ is an unimportant infinit constant and $\L$ is given by (\ref{lagr}).

Next we introduce the thermodynamical potential, $\Omega$, difined by

\begin{equation}
\label{termpot}
\Omega (T, \nu)=- \frac{T}{V}\ln Z
\end{equation}
where $lnZ= \frac {V}{T}(U(\nu)+ D_{1,2,3} m_{\psi,\pi,\sigma}^4)+lnZ_0+lnZ_I$. Since the interaction action $S_{I}$ contains terms which are more than quadratic in the fields, it is not possible to carry out the functional integration above in closed form. For a while we will neglect $S_{I}$ in our calculations. This amounts to considering only the tadpole contributions. Thus,

\begin{eqnarray}
\label{potef1}
\Omega_{1} (T, \nu) \equiv \frac{\lambda}{4}(f_{\pi}^2-\nu^2)^2 -c\nu - \left[ D_{1} m_{\psi}^4 + D_{2} m_{\pi}^4 + D_{3} m_{\sigma}^4 \right]
 - \frac{T}{V}lnZ_{0}
=\\ \frac{\lambda}{4}(f_{\pi}^2-\nu^2)^2 -c\nu - \left[ D_{1} m_{\psi}^4 + D_{2} m_{\pi}^4 + D_{3} m_{\sigma}^4 \right]
\nonumber
+ \\ \int \frac {d^3 p}{(2 \pi)^3}\left\{ \frac{1}{2} \omega_{\sigma} +T \ln (1-e^{-\beta \omega_{\sigma}})+\frac{3}{2} \omega_{\pi} +3T \ln (1-e^{-\beta \omega_{\pi}})-2 \cdot 2\left[ \omega_{\psi} +2T \ln (1+e^{-\beta \omega_{\psi}})\right] \right\}
\nonumber
\end{eqnarray}
with $\omega_{\sigma}^2 \equiv \bp^2 + m_{\sigma}^2$, $\omega_{\pi}^2 \equiv \bp^2 + m_{\pi}^2$ and $\omega_{\psi}^2 \equiv \bp^2 + m_{\psi}^2$. In the 
third line of (\ref{potef1}) the first factor 2 multiplying the bracket which contains the fermion contribution, comes from the spin degrees of freedom, whereas the other factor two is due the isospin degrees of freedom. Inside 
this same bracket there is another factor 2 corresponding to the particle 
and antiparticle contributions. The thermodynamical potential $\Omega_{1}$ 
is precisely the one-loop effective potential \cite{Dolan,Ram-Mohan} of 
the linear sigma model, and it can be expressed as,

\begin{equation}
\label{potef2}
\Omega_{1} (T, \nu)= U(\nu)+ D_{1,2,3} m_{\psi,\pi,\sigma}^4 + \Omega_{1}^{0} (\nu)+ \Omega_{1}^{\beta} (T, \nu)
\end{equation}

The equation of state of the noninteracting system composed by a (free) relativistic boson and fermion gas is

\begin{eqnarray}
\label{press}
P_0= \frac{T}{V}\ln Z_0 = -\Omega_1^\beta (T, \nu)
= \\ -T \int \frac {d^3 p}{(2 \pi)^3} \left[ \ln (1-e^{-\beta \omega_{\sigma}})+3 \ln (1-e^{-\beta \omega_{\pi}}) -8 \ln (1+e^{-\beta \omega_{\psi}})\right]
\nonumber 
\end{eqnarray}
where $P_0$ is the thermal pressure.

The integration over the temperature independent terms $\Omega_{1}^{0} (\nu ) \equiv \int \frac {d^3p}{(2\pi)^3}[\frac {1}{2}\omega_{\sigma}+\frac {3}{2}\omega_{\pi}-4\omega_{\psi}]$ actually diverges. It is exactly the counterterms $D_{1,2,3} m_{\psi,\pi,\sigma}^4$ which will take care of these divergences. We discuss this further in section \ref{Ren}. It can be known, 
from thermodynamical considerations, that in thermal equilibrium $\Omega$ is 
a minimum with respect to variations in $\nu$\cite{Kapusta}. Applying this extremum condition, we have 

\begin{equation}
\frac{\partial \Omega_{1} (T,\nu)} {\partial \nu}=0
\label{derpot1}
\end{equation}
Since $\Omega_{1}^{\beta} (T=0, \nu)=0$, the divergent quantity 
$\frac{\partial \Omega_{1}^0 (\nu)}{\partial \nu}=\int \frac {d^3 p}{(2 \pi)^3} \left[3 \lambda(\frac {1}{2\omega_{\sigma}}+\frac {1}{2\omega_{\pi}})-8g^2 \frac{1}{2\omega_{\psi}}\right]\nu $ represents the sum of the tadpoles at $T=0$. On the other hand, $\frac{\partial \Omega_{1}^{\beta}(T,\nu)}{\partial \nu}$ is finite at $T \ne 0$ because of the natural cutoff $\frac {1}{(e^{\beta \omega_{\sigma \pi,(\psi)}}-(+)1)}$ present in the integrals. We could let the counterterms absorb the finite parts of $\frac{\partial \Omega_{1}^0 (\nu)} {\partial \nu}$ together with the infinities since the vacuum contribution is irrelevant for the discussions of thermodynamics. But, as we will see in 
section \ref{loop3}, these terms are important in the verification of the Goldstone theorem in our self-consistent treatment. After these 
considerations, equation (\ref{derpot1}) is written as

\begin{equation}
\label{meanf1}
\left[F\left(\omega_{\pi,\sigma,\psi} ( m_{\pi,\sigma,\psi}),T \right) + G\left(\omega_{\pi,\sigma,\psi}( m_{\pi,\sigma,\psi}) \right) -\lambda(f_{\pi}^2-\nu^2) \right]\nu  -c=0 
\nonumber
\end{equation}
where the functions $F$ and $G$ are defined as

\begin{equation}
\label{fT}
F\left(\omega_{\pi,\sigma,\psi} ( m_{\pi,\sigma,\psi}),T \right) \equiv
\int \frac {d^3 p}{(2 \pi)^3} \left[3\lambda \left(\frac {1}{\omega_{\sigma}(e^{\beta \omega_{\sigma}}-1)}+\frac {1}{\omega_{\pi}(e^{\beta \omega_{\pi}}-1)} \right)+8g^2 \frac{1}{\omega_{\psi} (e^{\beta \omega_{\psi}}+1)}\right] ,
\end{equation}

\begin{equation}
\label{g}
G\left(\omega_{\pi,\sigma,\psi}( m_{\pi,\sigma,\psi}) \right) \equiv
\frac{3\lambda}{(4\pi)^2}\left(m_\pi^2\ln (\frac{m_\pi^2}{\mu^2})+ 
m_\sigma^2\ln (\frac{m_\sigma^2}{\mu^2})\right)-\frac{8g^2}{(4\pi)^2} m_\psi^2\ln (\frac{m_\psi^2}{\mu^2}) ,
\end{equation}
respectively. In eq.(\ref{g}), $\mu$ is the renormalization scale.

As a first approximation, we consider only the thermal loop corrections to 
the effective potential. This approximation allows us to get an analytic expression for the approximate critical temperature.

\begin{equation}
\label{meanf2}
\int \frac {d^3 p}{(2 \pi)^3} \left[3\lambda \left(\frac {n_\pi(\omega_\pi)}{\omega_\pi}+\frac{n_\sigma(\omega_\sigma)}{\omega_\sigma} \right)+8g^2 \frac{n_\psi(\omega_\psi)}{\omega_\psi}\right]\nu -\lambda(f_{\pi}^2-\nu^2) \nu -c=0 
\end{equation}
where $n_{\pi,\sigma}$ and $n_{\psi}$ are the usual distribution functions 
for bosons and fermions given by
\begin{equation}
\label{f1}
n_{\pi,\sigma}(\omega_{\pi,\sigma};T) = \frac{1}{ e^{ \beta \omega_{\pi,\sigma}} - 1 } 
\end{equation}

\begin{equation}
\label{f2}
n_{\psi}(\omega_{\psi};T) = \frac{1}{ e^{ \beta \omega_{\psi}} + 1 } ,
\end{equation}
respectively.
In the above expression, when one minimizes the effective potential, one is summing the thermal tadpole contributions to the usually called mean field equation. The chiral condensate, $\nu$, which is a  non trivial solution of 
this integral equation now depends on $T$. This equation can be solved with an explicit analytic form in the high temperature limit. The leading terms in the high temperature approximation for this integral equation are

\begin{equation}
\nu^3 + \left[\frac{1}{2}(1+ \frac{2g^2}{3 \lambda})T^2 -f_{\pi}^2\right] \nu - \frac {c}{\lambda}=0.
\label{consol1}
\end{equation}
The above equation has a real solution that is a slowly decreasing function of temperature, but does not vanish. Thus, when $c \ne 0$ the symmetry is never restored. On the other hand, when $c=0$ the non trivial solution of (\ref{consol1}) is

\begin{equation}
\nu^2 = f_{\pi}^2 -\frac{1}{2}(1+ \frac{2g^2}{3 \lambda})T^2 
\label{consol2}
\end{equation}

The critical temperature is defined as the temperature where the condensate 
goes to zero. It is given by

\begin{equation}
T_c^2= \frac {2f_{\pi}^2} {(1+ \frac{2g^2}{3 \lambda})}
\label{tcrit1}
\end{equation}
It shows that the inclusion of fermions does not change the order of the phase transition, but only lowers the value of $T_c$. Note that the interactions of the mesons with the fermions forces the ``critical'' temperature to depend on the coupling constants $\lambda$ and $g$. If $g=0$ we recover the result of \cite{Bochkarev}.

\section{Inclusion of Loop Corrections}
\subsection{The first necessity: Beyond the mean field approximation}
\label{loop1}

Let us analyze the finite temperature behavior of the tree-level meson 
masses  (\ref{massp}) and (\ref{masss}) as functions of the thermal 
expectation value of the sigma field, $\nu(T)$. Since $\nu^2(T)$ 
decreases as $T$ increases and $m^2<0$, the particle masses becomes 
tachyonic. Another problem which arises is the fact that Goldstone's 
theorem is not satisfied in the ordered phase (when $c=0$), i.e., 
replacing (\ref{consol2}) on (\ref{massp}), we obtain a non zero pion 
mass given by 
$m_{\pi}^2 = - \frac{\lambda}{2}(1+ \frac{2g^2}{3 \lambda})T^2 $. 
This pathological behavior is due to the fact that in our approximation 
we have neglect the interaction action $S_{I}$ in the thermodynamical 
potential (\ref{termpot}). The result is that the mean field approximation 
can be trusted only in the approximate prediction of a phase transition at $T_{c}$ given by (\ref{tcrit1}). It is incorrect in what concern the 
description of the finite temperature behavior of the meson and fermion 
masses. So it is necessary to include all one-loop corrections from all 
1PI diagrams present in $S_{I}$ to the masses.

Following the program of \cite{Kapusta} we expand the partition function 
in powers of the interaction, in order to get the one-loop self-energy corrections.

\begin{equation}
\ln Z_I=\ln (1+ \sum_{n=1}^{\infty } \frac{1}{n!} \frac {\int [d\phi]e^{S_0} S_I^n}
{[d \phi]e^{S_0}})
\label{lnint1}
\end{equation}

The one-loop 1PI graphs come from $\ln Z_1 + \ln Z_2$, which are given by

\begin{equation}
\ln Z_1 = \frac {\int [d\phi]e^{S_0} S_I}{[d \phi]e^{S_0}} 
\label{lnint2_0}
\end{equation}

\begin{equation}
\ln Z_2 = - \frac{1}{2} (\frac {\int [d\phi]e^{S_0} S_I}{[d \phi]e^{S_0}})^2 + \frac{1}{2} \frac {\int [d\phi]e^{S_0} S_I^2}{[d \phi]e^{S_0}}
\label{lnint2_1}
\end{equation}
where the disconnected diagrams cancel in $lnZ_2$ and the diagrams which gives rise to tadpoles in the self-energy are not to be included, since their effect is already considered in the mean field equation. The terms which ``survive'' come from:

\begin{eqnarray}
\label{lnint3}
\ln Z_I^{2-loop}= - \frac {\lambda}{4} \frac {\int [d\phi]e^{S_0} \int d \tau \int d^3 x (\pi^2 + \sigma^2)^2}{\int [d\phi]e^{S_o}}
+ \\ \frac {1}{2}\int_0^{\beta} d \tau_1 d \tau_2 \int d^3 x_1 d^3 x_2 \frac {\int [d \phi]e^{S_0} \left\{ {\lambda^2 \nu^2} [(\sigma \pi^2)^2+(\sigma^3)^2] +  {g^2}[(\overline{\psi} \sigma \psi)^2 + (\overline{\psi}i \gamma^5{\vec\pi}\cdot {\vec\tau}\psi)^2]\right\}}{\int [d \phi]e^{S_0}}
\nonumber
\end{eqnarray}
The 1PI graphs from this expression can be represented diagrammatically as 
shown in Fig.\ref{one}.

The self-energy for bosons and fermions are defined, respectively by

\begin{equation}
\label{self1}
\bD(\omega_n, \bk)_{\sigma,\pi}^{-1}= \bD_{0 \sigma,\pi}(\omega_n, \bk) ^{-1}+ \Pi_{\sigma,\pi} (\omega_n, \bk)
\end{equation}

\begin{equation}
\label{self2}
{\cal D}(\omega_{n}, \bk)^{-1}=  {\cal D}_0(\omega_{n}, \bk)^{-1}+ \Sigma(\omega_n, \bk)
\end{equation}
where $\bD_{0 \sigma,\pi}(\omega_{n}, \bp)$ and ${\cal D}_0(\omega_{n}, \bp)$ are the tree-level boson and fermion propagators, expressed respectively as

\begin{equation}
\label{prop1}
\bD_{0 \sigma,\pi}(\omega_n, \bk) ^{-1}= \omega_{n}^2+ \bk^2 + m_{\sigma,\pi}^2
\end{equation}

\begin{equation}
\label{prop2}
{\cal D}_{0}( \omega_{n}, \bk)^{-1}= \gamma_{\mu} k^{\mu}- m_{\psi}
\end{equation}
Here, $\omega_{n}$ are the Matsubara frequencies, defined as $\omega_{n}=2n\pi T$ for bosons and $\omega_{n}=(2n+1)\pi T \equiv \omega_{n f}$ for fermions.

To one-loop order the self-energy expressions are given\cite{Kapusta} by

\begin{equation}
\label{self3}
\Pi_{\sigma,\pi}=-2 (\frac{\delta \ln Z_I^{2-loop}}{\delta {\bD}_{0 \sigma,\pi}})_{1PI}
\end{equation}

\begin{equation}
\label{self4}
\Sigma = (\frac{\delta \ln Z_I^{2-loop}}{\delta {\cal D}_{0 \psi}})_{1PI}
\end{equation}

The self-energy graphs to each particle can be pictorially represented as by cutting one of the corresponding loops in the diagrams representing $(\ln Z_I^{2-loop})_{1PI}$. After the integration in $x$, $\tau$ and in the fields 
in $ (\ln Z_I^{2-loop})_{1PI}$ and the differentiations above, we obtain the following expressions for the self-energies at one loop order

\begin{eqnarray}
\label{self5}
\Pi_{\pi}(k_0,\bk)=\sum_{i=1}^4 \Pi_{\pi i}= 
\lambda 5T \sum_{n} \int \frac {d^3 p}
{(2 \pi)^3}\bD_{0 \pi}(\omega_n, \bp)
+ \lambda T \sum_{n} \int \frac {d^3 p}
{(2 \pi)^3} \bD_{0 \sigma}(\omega_n, \bp)+ \\
- 4\lambda^2\nu^2T \sum_{n}\int \frac {d^3 p}{(2\pi)^3}\bD_{0 \sigma}(\omega_{n+l}, \bp+ \bk) \bD_{0 \pi}(\omega_n, \bp) + g^2T \sum_{n}\int \frac {d^3 p}{(2\pi)^3} \Tr[\gamma^5 {\cal D}_{0 \psi}(\omega_{n+l}, \bp+ \bk) \gamma^5 {\cal D}_{0 \psi}(\omega_n, \bp)]
\nonumber
\end{eqnarray}

{\begin{eqnarray}
\label{self6}
\Pi_{\sigma}(k_0,\bk)= \sum_{i=1}^5 \Pi_{\sigma i}= 3 \lambda T \sum_{n} \int \frac {d^3 p}{(2 \pi)^3} \bD_{0 \sigma}(\omega_n, \bp)
+ 3 \lambda T \sum_{n} \int \frac {d^3 p}
{(2 \pi)^3} \bD_{0 \pi}(\omega_n, \bp) + \\
- 6\lambda^2\nu^2T \sum_{n}\int \frac {d^3 p}{(2\pi)^3}\bD_{0 \pi}(\omega_{n+l}, \bp+ \bk) \bD_{0 \pi}(\omega_n, \bp)
\nonumber - 18\lambda^2\nu^2T \sum_{n}\int \frac {d^3 p}{(2\pi)^3}\bD_{0 \sigma}(\omega_{n+l}, \bp+ \bk) \bD_{0 \sigma}(\omega_n, \bp) + \\
g^2T \sum_{n}\int \frac {d^3 p}{(2\pi)^3} \Tr[{\cal D}_{0 \psi}(\omega_{n+l}, \bp+ \bk) {\cal D}_{0 \psi}(\omega_n, \bp)]
\nonumber
\end{eqnarray}

\begin{eqnarray}
\label{self7}
\Sigma(k_0,\bk)= \sum_{i=1}^2 \Sigma_{i}=- g^2T \sum_{n}\int \frac {d^3 p}{(2\pi)^3}\bD_{0 \sigma}(\omega_{n+l}, \bp+ \bk) {\cal D}_{0 \psi}(\omega_n, \bp) + \\
- 3g^2T \sum_{n}\int \frac {d^3 p}{(2\pi)^3}\bD_{0 \pi}(\omega_{n+l}, \bp+ \bk) {\cal D}_{0 \psi}(\omega_n, \bp)
\nonumber
\end{eqnarray}

The diagrams representing the pion, sigma and nucleon one-loop self-energies 
are drawn in figures \ref{two},\ref{three} and \ref{four}.

We note here that we could get the same results for the self-energies 
directly applying Feynman rules to construct the diagrams with the appropriate substitutions: the $\delta$-function at each vertex is replaced with a 
Kronecker delta which imposes conservation of the discrete energy ($k_0=i\omega_{n}$), and round each loop of a thermal graph with $\int \frac{d^4k}{(2\pi)^4} \to \frac {i}{\beta}\sum_{n}\int \frac{d^3k}{(2\pi)^3}$,which are the finite-temperature Feynman rules\cite{Kapusta,Landshoff}. When the summation over $n$ is performed, 
each graph of the self-energies is separated into two parts, namely a temperature independent part(at this stage), which is divergent, and a temperature dependent part containing the Bose-Einstein distribution in 
the case of bosons or the Fermi-Dirac distribution for fermions. 

We will adopt the definition of mass at finite temperature as the real 
part of the pole of the corrected propagator at zero momentum $(\bk=0)$. 
Thus, from eqs. (\ref{self1}) and (\ref{self2}) we have
 
\begin{eqnarray}
\label{p1}
\bD_{\pi}(\omega_n, \left|\bk\right|=0) ^{-1}= \omega_{n}^2 + m_{\pi}^2 +\Pi_{\pi} (\omega_n, \left|\bk\right|=0)=0 \to \\
-k_{0 ,\pi}^2+ m_{\pi}^2 + \Pi_{\pi} (T,k_{0 ,\pi},\left|\bk\right|=0)=0
\nonumber
\end{eqnarray}

\begin{eqnarray}
\label{p2}
\bD_{\sigma}(\omega_n, \left|\bk\right|=0) ^{-1}= \omega_{n}^2 + m_{\sigma}^2 +\Pi_{\sigma} (\omega_n, \left|\bk\right|=0)=0 \to \\ -k_{0 ,\sigma}^2+ m_{\sigma}^2 +\Pi_{\sigma}  (T,k_{0 ,\sigma}, \left|\bk\right|=0)=0
\nonumber
\end{eqnarray}

\begin{eqnarray}
\label{p3}
{\cal D}_{\psi}(\omega_nf, \left|\bk\right|=0)^{-1}=\gamma^{\mu}k_{\mu}-m_{\psi}+\Sigma_{s} +\gamma^{\mu}\Sigma_{\mu} =0 \to \\ k_{0 ,\psi}+ \Sigma_{0}(T,k_{0,\psi}, \left|\bk\right|=0)+ \Sigma_{s} (T,k_{0,\psi} \left|\bk\right|=0)-m_{\psi}=0,
\nonumber
\end{eqnarray}
where the arrow indicates an analytical continuation from discrete to 
continuous energies in Minkowski space. Hence the physical masses are the 
values of the $k_{0;\pi,\sigma,\psi}$ which are the zeros of the functions (\ref{p1}), (\ref{p2}) and (\ref{p3}) above, i.e., the location of the poles 
in the limit $\bk=0$.
The full self-energy expressions $\Pi_\pi$, $\Pi_\sigma$, $\Sigma_{0}$ and $\Sigma_{s}$ are shown explicitly in appendix \ref{apend2}. The renormalization of the self-energy is studied in section \ref{Ren}. Through out this paper, 
we will use dimensional regularization, but omitting, for notational simplicity the factor $\mu^{2\epsilon}$ which multiplies $\lambda$. Since our calculations does not require traces involving an odd number of $\gamma^5$ matrices, we use the definition of $\gamma^5$ as in\cite{Chanowitz,Arnold}.

Since these expressions are self-consistent they have to be solved numerically. 
For each fixed temperature one finds a value of $M$ which satisfies the equations above. 
On the other hand, if one is interested only in the meson sector of the linear sigma model, 
the integrals in the self-energies could be evaluated exactly in the high temperature limit 
and at low frequency where the boson diagrams involving three-point vertices which are 
proportional to $\lambda^2\nu^2$ may be neglected. 
This is not consistent if one wants to study the behavior of the condensate $\nu$ 
and the particle masses in all ranges of temperatures. 
It is important to note that when $c=0$ the three-point vertex boson diagrams are 
significant in the region $T_{i}(\sim 0)<T<T_{f} (\sim\nu(T))$ 
and when $c\ne 0$, $\nu(T) \ne 0$ for any finite value of T.

\subsection{The second necessity: The resummation}
\label{loop2}

The expressions for the self-energies appearing in eq. (\ref{p1}), (\ref{p2}) and (\ref{p3}) 
are functions of $\omega_{\pi,\sigma,\psi}$ which are expressed in terms of the 
mean-field masses. As we discussed, the meson masses become negative as the temperature 
increases. Thus, in the computation of the one-loop corrections, the masses running in the 
loops become tachyonic. A proper resummation of higher order loops is naturally 
necessary \cite{Dolan}. Various resummation methods have been proposed in a tentative 
of curing the problem of the breaking down of the perturbative expansion at high temperature. 
In effective models, when a phase transition occurs, one can find tachyonic masses even 
below $T_c$. The O(N) linear $\sigma$  model which is one of the laboratory effective 
models employed to study QCD has been investigated by different authors using different 
techniques. One of these methods is the CJT formalism \cite{Cornwall} which provides for a 
consistent loop expansion of the effective potential in terms of the full propagator. 
The CJT formalism elegantly provides for a gap equation from stationarity conditions for 
the daisy and super-daisy effective potential. However some authors use this non-perturbative 
approach with an ansatz for the full (corrected) propagator in which the thermal 
corrections are momentum independent. These corrections are the finite piece of the 
divergent integral (which is temperature dependent through the gap equation for $M$) 
plus the finite explicit temperature dependent piece. This is the Hartree approximation, 
and means resuming only the ``bubble diagrams'' that are dominant at high temperatures. 
Another non-perturbative approach widely found in the literature is the large-$N$ approximation. 
The $N \to \infty$ limit facilitates the calculations, but can lead to inaccuracies 
\cite{Camelia3,Petropoulos}. One must be careful in taking the large-$N$ limit since 
its truncation depends on the problem to be studied and the 
relevant value of $N$ \cite{Camelia3}. In this case the three-point vertex 
diagrams are omitted which in principle makes sense only in the $N \to \infty$ limit 
since these sunset diagrams are of order $1/N$. In these two kinds of 
treatment one can not study the bosons interacting with fermions (with the interactions of the linear sigma model) since the 
self-energy diagrams are momentum dependent which invalidates the ansatz cited above. 
It is worth to remember that the Hartree approximation does not satisfies 
the Goldstone theorem\cite{Petropoulos,Lenaghan}. This fact may 
be attributed to the non inclusion of these diagrams. 
Although the finite temperature mass in these approaches is 
the pole of the corrected propagator, it is not the true mass, 
since their $\Pi_{\pi,\sigma}$ are not the true one-loop 
self-energy functions (see discussion below). 
The corrections included only shift the masses.
Once we are interested in the study of the 
masses behavior also in the range $0<T<T_c$ (if $c=0$) the 
three-particle vertex diagrams will not be neglected. 
The inclusion of these diagrams brings an additional 
complication since the self-energy now depends on the momentum.

\subsection{A non-perturbative resummation method: The MSCR}
\label{loop3}

Let us now introduce our procedure which resumms higher loop diagrams in the mean-field (tree-level) propagators. The method consist in recalculating the 
self-energy, in steps, using in each step the masses obtained in the previous 
one such that $ M_n ^2= (A_n+1)M_{n-1} ^2 + \Pi(M_{n-1}) $, where $n$ is the order of the non-perturbative correction and $A_n$ is the coefficient of the appropriate counterterm. With this procedure it is easier to identify and absorb the divergent parts of the self-energy in order to have finite gap-equations. The goal is to make renormalization possible since the masses which multiply the divergences are necessarily the same as in counterterms. 

\begin{center}
{{\bf Application of MSCR}}
\end{center}

Here we apply the MSCR in the study of the chiral fermion meson model at finite temperature. The analysis of the problem has to be done carefully which will be divided into three regions.

\begin{center}
{{\bf Region I: The low temperature region}}
\end{center}

The first region is for $0 \leq T < T^*$, where $T^*$ 
is the temperature where $\nu(T^*)=f_\pi$. This implies that 
$m_\pi^2=0$ and consequently the appearance of infrared divergences in the 
self-energy. So, in this region

{\bf Step 1:}

Start with the mean-field effective Lagrangian where the condensate 
and the masses are given by:

\begin{equation}
\left[F\left(\omega_{\pi,\sigma,\psi} ( m_{\pi,\sigma,\psi;0}),T \right) + G\left(\omega_{\pi,\sigma,\psi}( m_{\pi,\sigma,\psi;0}) \right) -\lambda(f_{\pi}^2-\nu^2) \right]\nu  -c=0,
\label{loop3-0}
\end{equation}

\begin{equation}
M_{\pi,0}^2=m_{\pi}^2= m^2+\lambda\nu^2,
\label{loop3-1}
\end{equation}

\begin{equation}
M_{\sigma,0}^2=m_{\sigma}^2= m^2+3\lambda\nu^2,
\label{loop3-2}
\end{equation}

\begin{equation}
M_{\psi,0}=m_{\psi}=g\nu.
\label{loop3-3}
\end{equation}

{\bf Step 2:}

Evaluate the one-loop self-energy corrections to these masses from 
the equations presented in appendix \ref{apend2} and define the 
condensate and the first order corrected masses as

\begin{equation}
\left[F\left(\omega_{\pi,\sigma,\psi} ( M_{\pi,\sigma,\psi,0}),T \right) + G\left(\omega_{\pi,\sigma,\psi}( M_{\pi,\sigma,\psi,0}) \right) -\lambda(f_{\pi}^2-\nu^2) \right]\nu  -c=0,
\label{loop3-30}
\end{equation}

\begin{eqnarray}
M_{\pi,1}^2= M_{\pi,0}^2+ \Pi_\pi(M_{\pi,0}^2, M_{\sigma,0}^2, M_{\psi,0})= \\ 
\nonumber
(A_1+E_1+1)m^2+(\overline A_1+E_1+1)\lambda\nu^2 + \Pi_\pi(M_{\pi,0}^2, M_{\sigma,0}^2, M_{\psi,0})=\\
m^2+ \lambda\nu^2 + \Pi_{\pi}^{Ren}(M_{\pi,0}^2, M_{\sigma,0}^2, M_{\psi,0}),
\label{loop3-4}
\nonumber
\end{eqnarray}

\begin{eqnarray}
M_{\sigma,1}^2= M_{\sigma,0}^2+ \Pi_\sigma(M_{\pi,0}^2, M_{\sigma,0}^2, M_{\psi,0}) = \\ 
\nonumber
(B_1+F_1+1)m^2+(F_1 + 1)\lambda\nu^2 +(\overline B_1+1)2\lambda\nu^2+ \Pi_\sigma(M_{\pi,0}^2, M_{\sigma,0}^2, M_{\psi,0})=\\
m^2+ 3\lambda\nu^2 + \Pi_{\sigma}^{Ren}(M_{\pi,0}^2, M_{\sigma,0}^2, M_{\psi,0}),
\label{loop3-5}
\nonumber
\end{eqnarray}

\begin{eqnarray}
M_{\psi,1}= M_{\psi,0}- \Sigma(M_{\pi,0}^2, M_{\sigma,0}^2, M_{\psi,0})=\\
\nonumber
(C_1+1)g\nu - \Sigma(M_{\pi,0}^2, M_{\sigma,0}^2, M_{\psi,0})=\\
g\nu - \Sigma^{Ren}(M_{\pi,0}^2, M_{\sigma,0}^2, M_{\psi,0}).
\label{loop3-6}
\nonumber
\end{eqnarray}
where $\Pi_{\pi,\sigma}=\Pi_{\pi,\sigma}(k_{0 ,\pi,\sigma}=M_{\pi,\sigma};\bk=0)^0+ \Pi_{\pi,\sigma} (T; k_{0 ,\pi,\sigma}=0;\bk=0)^{\beta}$, $A_1$, $\overline A_1$, $B_1$, $\overline B_1$, $C_1$, $E_1$ and $F_1$ are the appropriate coefficients of the counterterms added to the mean-field effective Lagrangian needed to render the model finite up to this order, which are shown in section \ref{Ren}. The requirement that $k_{0,\pi,\sigma}=0$ in $\Pi_{\pi,\sigma}^{\beta}$ excludes the possibility of tachyonic tree-level masses since thermal effects provides for the pion a 
non-zero width due to the Landau damping process. For the fermions we adopt 
the requirement that 
$\Sigma = \Sigma (k_{0,\psi}=M_{\psi};\bk=0)^0+ \Sigma (T; k_{0,\psi}=0;\bk=0)^{\beta}$ to prevent a similar consequence. 
The resummation has to be done exactly to avoid this problem. 
In this range there is no necessity of resummation since the masses running in 
the loops are positive, i.e., $M_{\pi,0}^2>0$, $M_{\sigma,0}^2>0$ and $M_{\psi,0}>0$.
\\
{\bf Renormalization:}
\\
The renormalization is done normally since the masses multiplying the divergences are the same as in the counterterms. The coefficients of the counterterms are found in section VI.
\\
{\bf Goldstone's Theorem:}

In the exact chiral limit ($c=0$) and low temperature phase (where $\nu \ne 0$) from eq. (\ref{p1}) at $( k_0 \to 0, \left|\bk\right|=0)$, we have

\begin{eqnarray} \label{tg0}
M_{\pi,1}^2=M_{\pi,0} ^2+ \Pi_\pi(k_0 \to 0, \left|\bk\right|=0)=
m^2 +\lambda\nu^2 +\Pi_{\pi}^{Ren} (k_0 \to 0, \left|\bk\right|=0)=\\
\nonumber
- \left\{\frac{3\lambda}{(4\pi)^2}\left[M_{\pi,0}^2\ln \left(\frac{M_{\pi,0}^2}{e\mu^2}\right)+ M_{\sigma,0}^2\ln \left(\frac{M_{\sigma,0}^2}{e\mu^2}\right)\right]-\frac{8g^2}{(4\pi)^2} M_{\psi,0}^2\ln \left(\frac{M_{\psi,0}^2}{e\mu^2}\right) + \right. \\
\nonumber
\left. \int_0^\infty\frac{dp p^2}{2\pi^2} \left[ 3\lambda \left( \frac{n_\sigma(M_{\sigma,0})}{\omega_\sigma(M_{\sigma,0})}+ \frac{n_\pi(M_{\pi,0})}{\omega_\pi(M_{\pi,0})}\right)+8g^2\frac{n_\psi(M_{\psi,0})}{\omega_\psi(M_{\psi,0})} \right] \right\} +\\
\nonumber
\frac{5\lambda}{(4\pi)^2}M_{\pi,0} ^2\ln \left(\frac{M_{\pi,0} ^2}{e\mu^2} \right)+\frac{5\lambda}{2}\int_0^{\infty }\frac{dp p^2}{\pi^2} \frac {n_{\pi}}{\omega_{\pi}}+\frac{\lambda}{(4\pi)^2}M_{\sigma,0} ^2\ln \left(\frac{M_{\sigma,0} ^2}{e\mu^2} \right)+ \frac{\lambda}{2}\int_0^{\infty }\frac{dp p^2}{\pi^2}\frac {n_{\sigma}}{\omega_{\sigma}}+\\
\nonumber
-4\frac{\lambda^2 \nu^2}{(4\pi)^2}\frac{ M_{\pi,0} ^2\ln \left(\frac{M_{\pi,0} ^2}{e\mu^2} \right)- M_{\sigma,0} ^2\ln \left(\frac{M_{\sigma,0} ^2}{e\mu^2} \right)}{ M_{\sigma,0}^2- M_{\pi,0}^2}
-2 \lambda^2 \nu^2 \int_0^{\infty}\frac{dp p^2}{\pi^2}\left[\frac {n_{\pi}}{\omega_{\pi}} \frac{1}{ M_{\sigma,0}^2- M_{\pi,0}^2} - \frac {n_{\sigma}}{\omega_{\sigma}} \frac{1}{ M_{\sigma,0}^2- M_{\pi,0}^2}\right]+\\
\nonumber
-\frac{8g^2}{(4\pi)^2}M_{\psi,0} ^2 \ln \left(\frac{M_{\psi,0} ^2}{e\mu ^2} \right)+
4g^2\int_0^{\infty}\frac{dp p^2}{\pi^2}\frac {n_{\psi}}{\omega_{\psi}}=0
\nonumber
\end{eqnarray} since $ M_{\sigma,0}^2- M_{\pi,0}^2=2\lambda\nu^2$ (in deriving eq.(\ref{tg0}) we have used eq.(\ref{meanf1})).

\begin{center}
{{\bf Region II: The intermediate temperature region}}
\end{center}

This is the region for $ T^* \leq T \leq T_c $ where the resummation is really necessary, since the masses in the loops are zero or tachyonic. The problem here 
is more complicated than in the other temperature regions since phase transition 
takes place. As is well known, around the critical temperature quantum fluctuations become essential and one loop corrections may not be enough. In fact, as we proceed to show here, this scheme (i.e., allowing only one 
loop corrections in the perturbative expansion) prevents us from adequately renormalizing the gap equations for the tree-level resummed masses and 
satisfying Goldstone's theorem. 

We show next that this is indeed the case and that a possible (but inconsistent !) way of achieving renormalization, would be e.g. to consider only the diagrams 
for which $\Pi_\pi=\Pi_\sigma$. This is very frequently implemented in the 
literature \cite{Chiku,Roh1,Petropoulos,Lenaghan}.

We will adopt the point of view that the MSCR fails in this region and higher 
order loop corrections in the perturbative expansion will be necessary at this stage. This is subject of current investigation. However we explicitly show 
were the problem appears.

{\bf Step 1:}

Start with the mean-field effective Lagrangian where the condensate 
and the masses are given by:

\begin{equation}
\left[F\left(\omega_{\pi,\sigma,\psi} ( m_{\pi,\sigma,\psi;0}),T \right) + G\left(\omega_{\pi,\sigma,\psi}( m_{\pi,\sigma,\psi;0}) \right) -\lambda(f_{\pi}^2-\nu^2) \right]\nu  -c=0,
\label{loop3-01}
\end{equation}

\begin{equation}
M_{\pi,0}^2=m_{\pi}^2= m^2+\lambda\nu^2,
\label{loop3-02}
\end{equation}

\begin{equation}
M_{\sigma,0}^2=m_{\sigma}^2= m^2+3\lambda\nu^2,
\label{loop3-03}
\end{equation}

\begin{equation}
M_{\psi,0}=m_{\psi}=g\nu.
\label{loop3-04}
\end{equation}

{\bf Step 2:}

Evaluate the one-loop self-energy corrections to these masses from 
the equations presented in appendix \ref{apend2} and define the 
condensate and the first order corrected masses as

\begin{equation}
\left[F\left(\omega_{\pi,\sigma,\psi} ( M_{\pi,\sigma,\psi,0}),T \right) + G\left(\omega_{\pi,\sigma,\psi}( M_{\pi,\sigma,\psi,0}) \right) -\lambda(f_{\pi}^2-\nu^2) \right]\nu  -c=0,
\label{loop3-05}
\end{equation}

\begin{eqnarray}
M_{\pi,1}^2= M_{\pi,0}^2+ \Pi_\pi(M_{\pi,0}^2, M_{\sigma,0}^2, M_{\psi,0})= \\ 
\nonumber
(A_1+E_1+1)m^2+(\overline A_1+E_1+1)\lambda\nu^2 + \Pi_\pi(M_{\pi,0}^2, M_{\sigma,0}^2, M_{\psi,0})=\\
m^2+ \lambda\nu^2 + \Pi_{\pi}^{Ren}(M_{\pi,0}^2, M_{\sigma,0}^2, M_{\psi,0}),
\label{loop3-06}
\nonumber
\end{eqnarray}

\begin{eqnarray}
M_{\sigma,1}^2= M_{\sigma,0}^2+ \Pi_\sigma(M_{\pi,0}^2, M_{\sigma,0}^2, M_{\psi,0}) = \\ 
\nonumber
(B_1+F_1+1)m^2+(F_1 + 1)\lambda\nu^2 +(\overline B_1+1)2\lambda\nu^2+ \Pi_\sigma(M_{\pi,0}^2, M_{\sigma,0}^2, M_{\psi,0})=\\
m^2+ 3\lambda\nu^2 + \Pi_{\sigma}^{Ren}(M_{\pi,0}^2, M_{\sigma,0}^2, M_{\psi,0}),
\label{loop3-07}
\nonumber
\end{eqnarray}

\begin{eqnarray}
M_{\psi,1}= M_{\psi,0}- \Sigma(M_{\pi,0}^2, M_{\sigma,0}^2, M_{\psi,0})=\\
\nonumber
(C_1+1)g\nu - \Sigma(M_{\pi,0}^2, M_{\sigma,0}^2, M_{\psi,0})=\\
g\nu - \Sigma^{Ren}(M_{\pi,0}^2, M_{\sigma,0}^2, M_{\psi,0}).
\label{loop3-08}
\nonumber
\end{eqnarray}
At this stage of the application of the method, the renormalization is possible,  Goldstone's theorem is 
verified, but the tree-level masses are zero or tachyonic.

{\bf Step 3:}

Now we take the masses computed in the previous step and improve the results defining a next-order non-perturbative correction. With this we get a new effective Lagrangian where the condensate and masses are given by

\begin{equation}
\left[F\left(\omega_{\pi,\sigma,\psi} ( M_{\pi,\sigma,\psi,1}),T \right) + G\left(\omega_{\pi,\sigma,\psi}( M_{\pi,\sigma,\psi,1}) \right) -\lambda(f_{\pi}^2-\nu^2) \right]\nu  -c=0,
\label{loop3-31}
\end{equation}

\begin{eqnarray}
\label{loop3-7}
M_{\pi,2}^2= M_{\pi,1}^2+ \Pi_\pi(M_{\pi,1}^2, M_{\sigma,1}^2, M_{\psi,1})=\\
\nonumber
(A_2+E_2+1)m^2+(\overline A_2+E_2+1)\lambda\nu^2 + \\ 
\nonumber
(\overline{\overline A}_2 +E_2+1) \Pi_{\pi}^{Ren}(M_{\pi,0}^2, M_{\sigma,0}^2, M_{\psi,0})+\Pi_\pi(M_{\pi,1}^2, M_{\sigma,1}^2, M_{\psi,1})= \\
m^2+ \lambda\nu^2 + \Pi_{\pi}^{Ren}(M_{\pi,1}^2, M_{\sigma,1}^2, M_{\psi,1}) -
\frac{\lambda }{(4\pi)^2}\Pi_{\sigma}^{Ren}(M_{\pi,1}^2, M_{\sigma,1}^2, M_{\psi,1})\frac{1}{\widetilde \epsilon},
\nonumber
\end{eqnarray}

\begin{eqnarray}
\label{loop3-8}
M_{\sigma,2}^2= M_{\sigma,1}^2+ \Pi_\sigma(M_{\pi,1}^2, M_{\sigma,1}^2, M_{\psi,1})=\\
\nonumber
(B_2+F_2+1)m^2+(F_2 + 1)\lambda\nu^2+(\overline B_2+1)2\lambda\nu^2 + \\ 
\nonumber
(\overline{\overline B}_2 +F_2+1) \Pi_{\sigma}^{Ren}(M_{\pi,0}^2, M_{\sigma,0}^2, M_{\psi,0})+ \Pi_\sigma(M_{\pi,1}^2, M_{\sigma,1}^2, M_{\psi,1}) = \\
m^2+ 3\lambda\nu^2 + \Pi_{\sigma}^{Ren}(M_{\pi,1}^2, M_{\sigma,1}^2, M_{\psi,1})-
\frac{3\lambda}{(4\pi)^2}\Pi_{\pi}^{Ren}(M_{\pi,1}^2, M_{\sigma,1}^2, M_{\psi,1})\frac{1}{\widetilde \epsilon},
\nonumber
\end{eqnarray}

\begin{eqnarray}
\label{loop3-9}
M_{\psi,2}= M_{\psi,1}- \Sigma(M_{\pi,1}^2, M_{\sigma,1}^2, M_{\psi,1})=\\
\nonumber
(C_2+1)g\nu -(\overline C_2+1)\Sigma^{Ren}(M_{\pi,0}^2, M_{\sigma,0}^2,M_{\psi,o}) - \Sigma(M_{\pi,1}^2, M_{\sigma,1}^2, M_{\psi,1})
= \\ g\nu - \Sigma^{Ren}(M_{\pi,1}^2, M_{\sigma,1}^2, M_{\psi,1}).
\nonumber
\end{eqnarray}

It is shown that in the $\lambda \phi ^4$ model, this (first) 
recalculation is equivalent to the sum of an infinite set of diagrams, 
namely the ``daisy'' sum \cite{Dolan} or the set of ring \cite{Kapusta}. 
The coefficients of the temperature dependent mass counterterms $\overline{\overline A}_2$, $\overline{\overline B}_2$ and $\overline C_2$ 
are fixed in a manner to cancel not only divergences proportional to $\Pi_{\pi}^{Ren}(M_{\pi,0}^2, M_{\sigma,0}^2, M_{\psi,0})$, $\Pi_{\sigma}^{Ren}(M_{\pi,0}^2, M_{\sigma,0}^2, M_{\psi,0})$  and $\Sigma^{Ren}(M_{\pi,0}^2, M_{\sigma,0}^2,M_{\psi,o})$ respectively, 
but also these terms together. That is, at each stage of the procedure, 
for $n > 1$, in the expressions for $M_{\pi,\sigma,\psi,n}$, 
the self-energy $\Pi_{\pi,\sigma}(M_{\pi,\sigma,\psi,n-2})$ 
(or $\Sigma(M_{\pi,\sigma,\psi,n-2})$) 
have to be cancelled to avoid overcounting of diagrams. 

This shows explicitly that renormalization can not be performed within this approximation scheme. This is not surprising since in this temperature region quantum fluctuation may need a more thorough description. So, there is no reason to believe that only the ``daisy'' diagrams should be resummed at low and intermediate temperatures. In fact, the ``daisy'' graphs contributions are dominant at high temperature \cite{Dolan}.

{\bf Renormalization:}
\\
Since in this region $ M_{\sigma,n}^2- M_{\pi,n}^2=2\lambda\nu^2 + \Delta\Pi$, 
where $\Delta\Pi \equiv \Pi_{\sigma}-\Pi_{\pi}$, there is the presence of 
undesirable non-renormalizible terms. These terms are the last ones on the r.h.s. of equations (\ref{loop3-7}) and (\ref{loop3-8}) which come from equations 
(\ref{pio2}) and (\ref{sig2}) respectively and can not be absorbed in the 
counterterms.

{\bf Goldstone's Theorem:}
\\
In this region, Goldstone's theorem is satisfied only if $\Delta \Pi \to 0$ 
and the contribution in (\ref{pio3}) is decoupled into integrals proportional 
to $\lambda$. This would assure the cancellation of $M_\pi$ at $( k_0 \to 0, \left|\bk\right|=0)$. The reason for this 
frustration is the same as for the lack of renormalizability. 

{\bf Step 4:} (applicable only in the case where $\Delta\Pi=0$. This guarantees that renormalization and Goldstone theorem can be satisfactorily implemented at each step. In particular this will be the case for the high temperature region 
as will be shown next.)

Proceeding with the iteration, in the limit $n \to \infty$ the masses $M_n$ 
have formally the same expressions as the masses $M_{n-1}$ which are already renormalized. Thus, in this limit we will have, 

\begin{equation}
\left[F\left(\omega_{\pi,\sigma,\psi} ( M_{\pi,\sigma,\psi,n}),T \right) + G\left(\omega_{\pi,\sigma,\psi}( M_{\pi,\sigma,\psi,n}) \right) -\lambda(f_{\pi}^2-\nu^2) \right]\nu  -c=0,
\label{loop3-32}
\end{equation}

\begin{equation}
\label{loop3-10}
M_{\pi,n}^2=m^2+\lambda\nu^2 + \Pi_{\pi}^{Ren}(M_{\pi,n}^2, M_{\sigma,n}^2, M_{\psi,n}),
\end{equation}

\begin{equation}
\label{loop3-11}
M_{\sigma,n}^2=m^2+3\lambda\nu^2 + \Pi_{\pi}^{Ren}(M_{\pi,n}^2, M_{\sigma,n}^2, M_{\psi,n}),
\end{equation}

\begin{equation}
\label{loop3-12}
M_{\psi,n}=g\nu - \Sigma^{Ren}(M_{\pi,n}^2, M_{\sigma,n}^2,M_{\psi,n}).
\end{equation}
At each intermediate step, in the loops we set $k_{0 \pi, \psi}^2=M_{\pi,\psi,n-1}^2$ in the computation of $M_{\pi,\sigma,\psi,n}^2$. This ensures the cancellation of the divergences in all stages of the process, since the 
masses in the counterterms will necessarily be the same as in the divergences. In the end, in the resulting equations of interest (to be solved nummericaly), $K_{0 \pi, \psi}^2=M_{\pi,\psi}^2$ as it should. 
By our MSCR we have gotten a set of four coupled non-linear integral equations to be solved self-consistently, with finite gap equations for the tree-level masses, which read 

\begin{equation}
\label{p10}
M_{\pi}^2= m^2 +\lambda\nu^2 +\Pi_{\pi}^{Ren} (k_{0\pi}=M_\pi, \left|\bk\right|=0)
\end{equation}

\begin{equation}
\label{p11}
M_{\sigma}^2= m^2 +3\lambda\nu^2 +\Pi_{\pi}^{Ren} ( k_{0\pi}=M_\pi,\left|\bk\right|=0)
\end{equation}

\begin{equation}
\label{p12}
M_{\psi}= g\nu - \Sigma_{0}^{Ren}( k_{0\psi}=M_\psi,\left|\bk\right|=0)- \Sigma_{s}^{Ren}( k_{0\psi}=M_\psi,\left|\bk\right|=0)
\end{equation}

\begin{eqnarray}
\label{p13}
\nu \left\{ m^2+\lambda\nu^2 +\frac{3\lambda}{(4\pi)^2}\left[M_\pi^2\ln \left(\frac{M_\pi^2}{e\mu^2}\right)+ M_\sigma^2\ln \left(\frac{M_\sigma^2}{e\mu^2}\right)\right]-\frac{8g^2}{(4\pi)^2} M_\psi^2\ln \left(\frac{M_\psi^2}{e\mu^2}\right) + \right. \\ 
\left. \int_0^\infty\frac{dp p^2}{2\pi^2} \left[ 3\lambda \left( \frac{n_\sigma(M_\sigma)}{\omega_\sigma(M_\sigma)}+ \frac{n_\pi(M_\pi)}{\omega_\pi(M_\pi)}\right)+8g^2\frac{n_\psi(M_\psi)}{\omega_\psi(M_\psi)} \right] \right\}=c
\nonumber
\end{eqnarray}
where the renormalization scale $\mu$ can be determined by a physical condition. We choose $\mu$ such that the pion mass has the correct value at $T=0$.

Now we have to go back to real world encoded by eq. (\ref{p1} to \ref{p3}). 
In order to get finite physical masses from the pole of these equations it is necessary to sum and subtract the finite quantities $\Pi_\pi^{Ren}$ and $\Sigma^{Ren}$ which will be regarded as mass parameters \cite{Weinberg,Mallik,Parwani,Chiku}. This corresponds to the reorganization of 
the perturbative expansion. Now we rewrite eq. (\ref{lagr}) as

\begin{eqnarray}
{\cal L}  = - \frac{\lambda}{4}(f_{\pi}^2- \nu^2)^2 +c\nu + \overline{\psi}[i \gamma^{\mu} \partial_{\mu}  -M_{\psi}] \psi 
\nonumber
+ \\ 
\frac{1}{2} [(\partial \vec \pi)^2 -M_{\pi}^2 \vec \pi ^2 + (\partial \sigma)^2 -M_{\sigma}^2 \sigma ^2] -g \overline{\psi}[ \sigma +i \gamma^5{\vec\pi}\cdot {\vec\tau})]\psi +
\nonumber
 \\ 
- \frac{\lambda}{4}[(\vec\pi^2 + \sigma^2)^2 +4 \nu \sigma (\vec\pi^2 + \sigma^2)] + \Sigma^{Ren} \overline{\psi}\psi + \frac{1}{2}\Pi_{\pi}^{Ren} (\vec\pi^2 + \sigma^2)+ C.T.
\label{lagRES}
\end{eqnarray}
The last two terms on the third line of eq.(\ref{lagRES}) must be considered 
as extra ``interaction'' terms and will naturally be present in $\ln Z_I$, eq.(\ref{lnint3}). The counterterm structure of eq.(\ref{lagRES}) is the same 
as the one present in the method for the pion and fermion, differing only by numerical factors in the case of the sigma mass renormalization. By eqs.(\ref{self3}) and (\ref{self4}) the extra contribution to the self-energy read 
\begin{equation}
\label{extra1}
\Pi_{\pi}^{extra}=\Pi_{\sigma}^{extra}=-\Pi_{\pi}^{Ren}
\end{equation}

\begin{equation}
\label{extra2}
\Sigma^{extra}=\Sigma^{Ren}
\end{equation}
where we have defined
\begin{equation}
\label{extra3}
\Pi_{\pi}^{Ren} \equiv \Pi_\pi(k_{0 ,\pi}=M_{\pi},\bk=0)^0+ \Pi_{\pi} (T,k_0=0,\bk=0)^{\beta}
\end{equation}

\begin{equation}
\label{extra4}
\Sigma^{Ren} \equiv \Sigma (k_{0}=M_{\psi},\bk=0)^0+ \Sigma (T,k_0=0,\bk=0)^{\beta}
\end{equation}
As a result, the final resummmed tree-level masses (eqs.(\ref{p10}), (\ref{p11}) and (\ref{p12})), may be used in eqs. (\ref{p1}), (\ref{p2}) and (\ref{p3}) if one wants to study for instance spectral functions as the authors of\cite{Chiku}, or decay width as in \cite{Lenaghan}.

{\bf Goldstone's Theorem:}
\\
If the sunset type graphs are neglected, the self-energy function at one-loop is not complete, despite of the fact that the definition of masses as poles of propagators at zero momentum is still valid. Now, having the result for the full one-loop (and higher-order loops contributions from the resumation) self-energy function we can test algebraically the fulfillment of Goldstone's theorem in the exact chiral limit ($c=0$) and low temperature phase (where $\nu \ne 0$). From eq. (\ref{p1}) at $( k_0 \to 0, \left|\bk\right|=0)$, we have

\begin{eqnarray} \label{tg1}
M_\pi ^2 + \Pi_{\pi}^{total} = M_\pi ^2+ \Pi_\pi(k_0 \to 0, \left|\bk\right|=0) + \Pi_{\pi}^{extra}=m^2 +\lambda\nu^2 +\Pi_{\pi}^{Ren} (k_0 \to 0, \left|\bk\right|=0)=\\
\nonumber
- \left\{\frac{3\lambda}{(4\pi)^2}\left[M_\pi^2\ln \left(\frac{M_\pi^2}{e\mu^2}\right)+ M_\sigma^2\ln \left(\frac{M_\sigma^2}{e\mu^2}\right)\right]-\frac{8g^2}{(4\pi)^2} M_\psi^2\ln \left(\frac{M_\psi^2}{e\mu^2}\right) + \right. \\
\nonumber
\left. \int_0^\infty\frac{dp p^2}{2\pi^2} \left[ 3\lambda \left( \frac{n_\sigma(M_\sigma)}{\omega_\sigma(M_\sigma)}+ \frac{n_\pi(M_\pi)}{\omega_\pi(M_\pi)}\right)+8g^2\frac{n_\psi(M_\psi)}{\omega_\psi(M_\psi)} \right] \right\} +\\
\nonumber
\frac{5\lambda}{(4\pi)^2}M_\pi ^2\ln \left(\frac{M_\pi ^2}{e\mu^2} \right)+\frac{5\lambda}{2}\int_0^{\infty }\frac{dp p^2}{\pi^2} \frac {n_{\pi}}{\omega_{\pi}}+\frac{\lambda}{(4\pi)^2}M_\sigma ^2\ln \left(\frac{M_\sigma ^2}{e\mu^2} \right)+ \frac{\lambda}{2}\int_0^{\infty }\frac{dp p^2}{\pi^2}\frac {n_{\sigma}}{\omega_{\sigma}}+\\
\nonumber
-4\frac{\lambda^2 \nu^2}{(4\pi)^2}\frac{ M_\pi ^2\ln \left(\frac{M_\pi ^2}{e\mu^2} \right)- M_\sigma ^2\ln \left(\frac{M_\sigma ^2}{e\mu^2} \right)}{ M_{\sigma}^2- M_{\pi}^2}
-2 \lambda^2 \nu^2 \int_0^{\infty}\frac{dp p^2}{\pi^2}\left[\frac {n_{\pi}}{\omega_{\pi}} \frac{1}{ M_{\sigma}^2- M_{\pi}^2} - \frac {n_{\sigma}}{\omega_{\sigma}} \frac{1}{ M_{\sigma}^2- M_{\pi}^2}\right]+\\
\nonumber
-\frac{8g^2}{(4\pi)^2}M_\psi ^2 \ln \left(\frac{M_\psi ^2}{e\mu ^2} \right)+
4g^2\int_0^{\infty}\frac{dp p^2}{\pi^2}\frac {n_{\psi}}{\omega_{\psi}}=0
\nonumber
\end{eqnarray} since $ M_{\sigma}^2- M_{\pi}^2=2\lambda\nu^2$ (in deriving eq.(\ref{tg1}) we have used eq.(\ref{p13})).

To our opinion the fullfilment of Goldstone's theorem is ultimately related to the preservation of the relation imposed by chiral symmetry to the tree-level masses. Moreover, it is crucial to keep all diagrams of a given order. This is due to the fact that, strictly speaking, a loop expansion is an expansion in powers of the Lagrangian. As discussed in \cite{Kapusta} in order to respect the symmetries of the Lagrangian, one must retain all diagrams to the given number of loops. 

Then, in the absence of the explicit chiral symmetry breaking term, one has, \\
for $0<T<T_c$

\begin{eqnarray} \label{tg2}
M_{\pi}^2=0\\
\nonumber
M_{\sigma}^2=2\lambda\nu^2
\end{eqnarray}
for $T=T_c$
\begin{equation} \label{tg3}
M_{\pi}^2=M_{\sigma}^2=0
\end{equation}
and for $T>T_c$
\begin{equation} \label{tg4}
M_{\pi}^2=M_{\sigma}^2=m^2+3\lambda \int_0^{\infty}\frac{dp p^2}{\pi^2}\frac {n_{b}}{\omega_{b}}+4g^2\int_0^{\infty}\frac{dp p^2}{\pi^2}\frac {n_{f}}{\omega_{f}}
\end{equation}
which shows chiral symmetry restoration. Here b stands for bosons and f for fermions. We could interpret the result in the r.h.s. of eq. (\ref{tg4}) as if each independent pion effectively ``sees'' one sigma and the other two pions and four fermions (since $\b\mu$, the chemical potential, here is zero). On the other hand the sigma ``sees'' the three pions and four fermions. This equation serves to define the critical temperature in which the common masses of the particles vanish. In the high-temperature limit of these integrals, we find that $ T_c^2= \frac {2f_{\pi}^2} {(1+ \frac{2g^2}{3 \lambda})}$, as predicted by the mean-field analysis in eq.(\ref{tcrit1}).

\begin{center}
{{\bf Region III: The high temperature region }}
\end{center}

This is the region of high temperatures, $ T \geq T_c $, 
if $c=0$ and $\nu=0$ or $ T \geq T_i $, where $T_i$ is defined as a ``inflexion'' temperature, for the case $c \ne 0$ and $\nu << f_\pi$ such that 
$M_{\pi,0}^2 \approx M_{\sigma,0}^2=m^2$.

\begin{eqnarray}
M_{\pi,1}^2= M_{\pi,0}^2+ \Pi_\pi(M_{\pi,0}^2, M_{\sigma,0}^2, M_{\psi,0})= \\ 
\nonumber
m^2 + \Pi_\pi(M_{\pi,0}^2, M_{\sigma,0}^2, M_{\psi,0}),
\label{loop3-66}
\end{eqnarray}

\begin{eqnarray}
M_{\sigma,1}^2= M_{\sigma,0}^2+ \Pi_\sigma(M_{\pi,0}^2, M_{\sigma,0}^2, M_{\psi,0}) = \\ 
\nonumber
m^2 + \Pi_\sigma(M_{\pi,0}^2, M_{\sigma,0}^2, M_{\psi,0}) = \\
\nonumber
m^2 + \Pi_\pi(M_{\pi,0}^2, M_{\sigma,0}^2, M_{\psi,0}) = 
M_{\pi,1}^2 \equiv M_1^2,
\label{loop3-77}
\end{eqnarray}

\begin{eqnarray}
M_{\psi,1}= M_{\psi,0}- \Sigma(M_{\pi,0}^2, M_{\sigma,0}^2, M_{\psi,0})=\\
\nonumber
g\nu - \Sigma(M_{\pi,0}^2, M_{\sigma,0}^2, M_{\psi,0}).
\label{loop3-88}
\end{eqnarray}

Note that the pion and sigma masses become degenerate and the problem encountered in the previous region ($ T^* \leq T \leq T_c $) is no longer 
here since $\Delta\Pi=0$ in this region of temperatures. In this case, the masses in the loops can be neglected, and we have

\begin{eqnarray}
\label{lsig-7}
M_1 ^2= (A_1+1)M_0 ^2 + \Pi(M_0) = m^2 + \frac{\lambda}{2}\left(1+\frac{2g^2}{3\lambda} \right)T^2 =\\
\lambda f_\pi ^2 \left[ \frac{T^2}{T_c ^2} - 1 \right]
\nonumber
\end{eqnarray}
If we set $g=0$ these results agrees with the ones obtained by Bochkarev and Kapusta\cite{Bochkarev}.

Following the iterations, we find for the $n$-th iterated mass

\begin{eqnarray}
\label{lsig-8}
M_n ^2= (A_n+1)M_{n-1} ^2 + \Pi(M_{n-1}) = \\
m^2 + \frac{\lambda}{2}\left(1+\frac{2g^2}{3\lambda} \right)T^2 \left[1-\frac{3}{\pi T}M_{n-1} \right]
\nonumber
\end{eqnarray}

In the limit $n \to \infty$, we get

\begin{equation}
\label{lsig-9}
M ^2= m^2 + \frac{\lambda}{2}\left(1+\frac{2g^2}{3\lambda} \right)T^2 \left[1-\frac{3}{\pi T}M \right]
\end{equation}
which can be easily solved for $M$,

\begin{equation}
\label{lsig-10}
M=  \left[   \left(\frac{3\lambda f_\pi ^2}{2 \pi} \frac{T}{T_c ^2} \right)^2 +\lambda f_\pi ^2 \left(\frac{T^2}{T_c ^2}-1 \right) \right]^{\frac{1}{2}}-\frac{3\lambda f_\pi ^2}{2 \pi} \frac{T}{T_c ^2}.
\end{equation}
For $T>>T_c$

\begin{equation}
\label{lsig-11}
M ^2=\frac{\lambda}{2}\left(1+\frac{2g^2}{3\lambda} \right)T^2
\end{equation}

\section{The Massless $\lambda \phi^4$ At High-Temperature}

Now we apply the MSCR to study a very popular model: the massless $\lambda \phi^4$ model in the weak coupling limit 

\begin{equation}
\label{phi4-1}
{\cal L} = \frac{1}{2} (\partial_{\mu} \phi)^2 - \frac{\lambda}{4!}\phi^4
\end{equation}

\begin{eqnarray}
\label{phi4-2}
M_0=0 \\
M_1 ^2 = M_0 ^2 + \Pi(M_0)= \frac{\lambda T^2}{24}
\nonumber
\end{eqnarray}
at this stage of the procedure there is no necessity of adding counterterms since up to this order there are no ultraviolet divergences in dimensional regularization\cite{Parwani}. Here $\Pi$ is the 1PI one-loop self-energy to lowest order, namely the ``bubble'' of Fig. \ref{five}(a).

\begin{eqnarray}
\label{phi4-3}
M_2 ^2 = M_1 ^2 + \Pi(M_1)= (A_2 + 1)\Pi^{Ren}(M_0)+ \Pi(M_1)=\\
\nonumber
\frac{\lambda T^2}{24} \left(1- \frac{3M_1}{\pi T} \right) + O(\lambda^2 \ln \lambda)=\\
\nonumber
M_1 ^2 \left[1- 3\left(\frac{\lambda}{24\pi^2}\right)^{\frac{1}{2}} \right] + O(\lambda^2 \ln \lambda)
\end{eqnarray}
with the result that this correction to the mass is of order $\lambda^{\frac{3}{2}}$, which is an signature of the non-perturbative resummation. The temperature dependent counterterm is fixed so as to cancel 
the divergence and avoid overcounting of diagrams, as explained before. So, $A_2=-1+\frac{\lambda}{2(4\pi)^2}\frac{1}{\epsilon}$. {\it The diagrams used, in a given number of loops, in any resummation method must be the same in all stages of the process. What changes is the masses running in the loops at each iteration. This is because one must keep the same fundamental theory in the recalculation of the self-energy.} The result shown in eq. (\ref{phi4-3}) is in agreement with the one obtained by Parwani's resummed perturbative expansion\cite{Parwani} (see eq.(2.12) of his paper). The second iteration corrected mass, $M_2$, which was obtained in our method evaluating Fig. \ref{five}(a) with $M_1$ in that loop 
can equivalently 
be achieved calculating the ``daisy'' sum, that is a summation of the infinite set of ``daisy'' diagrams of Fig. \ref{five}(b) with $M_0$ in the loops. In 
this case all ``daisy'' types diagrams are IR-divergent since $M_0=0$, but 
their sum is IR-finite\cite{Kapusta,Parwani}.

Continuing the iterations, we find for the next correction

\begin{eqnarray}
\label{phi4-4}
M_3 ^2 = M_2 ^2 + \Pi(M_2)=\\
\frac{\lambda T^2}{24} \left[1- \frac{3M_1}{\pi T}\left(1-\frac{3M_1}{\pi T} \right)^{\frac{1}{2}} \right].
\nonumber
\end{eqnarray}
When $\lambda << 1$ we get

\begin{equation}
\label{phi4-5}
M_3 ^2 = \frac{\lambda T^2}{24} \left[1- 3 \left(\frac{\lambda}{24\pi^2}\right)^{\frac{1}{2}} + \frac{9}{2}\left(\frac{\lambda}{24\pi^2}\right)\right]
\end{equation}
and for the n-th iteration, we obtain

\begin{equation}
\label{phi4-6}
M_n ^2 = \frac{\lambda T^2}{24} \left\{1+ \sum_{j=1}^{n} \frac{1}{2^{j-1}} \left[-3\left(\frac{\lambda}{24\pi^2}\right)^{\frac{1}{2}} \right]^j \right\}
\end{equation}
The ``superdaisy'' sum \cite{Dolan} corresponds to the limit $n \to \infty$ of equation (\ref{phi4-6}) and it can be summed up (for $\lambda << 1$) to give 

\begin{equation}
\label{phi4-7}
M ^2 = \frac{\lambda T^2}{24} \left[ \frac{1-\frac{3}{2}\left(\frac{\lambda}{24\pi^2}\right)^{\frac{1}{2}}} {1+\frac{3}{2} \left(\frac{\lambda}{24\pi^2}\right)^{\frac{1}{2}}} \right]
\end{equation}

\section{Renormalization}
\label{Ren}

\subsection{Determination of the counter-terms}
The divergences are regulated via dimensional rugularization. To renormalize the divergences, we use the Minimal Subtraction scheme where only the poles are eliminated by the appropriate counterterms. The first-order parameters of the temperature-dependent counterterms read

\begin{equation}
\label{Ren1}
A_1=\frac{6 \lambda}{(4 \pi)^2} \frac{1}{\widetilde \epsilon}, \overline{A}_1=\frac{12 \lambda}{(4 \pi)^2} \frac{1}{\widetilde \epsilon}, E_1=\frac{4 g^2}{(4 \pi)^2} \frac{1}{\widetilde \epsilon},
\end{equation} 
with $\frac{1}{\widetilde \epsilon} \equiv \frac{2}{4-d}-\gamma + \log(4\pi)$, where $\gamma$ is the Euler constant.

\begin{equation}
\label{Ren2}
B_1=\frac{6 \lambda}{(4 \pi)^2} \frac{1}{\widetilde \epsilon}, \overline{B}_1=\frac{6 \lambda}{(4 \pi)^2} \frac{1}{\widetilde \epsilon}, F_1=\frac{4 g^2}{(4 \pi)^2} \frac{1}{\widetilde \epsilon}
\end{equation}

\begin{equation}
\label{Ren3}
C_1=\frac{8 g^2}{(4 \pi)^2} \frac{1}{\widetilde \epsilon}
\end{equation}

\begin{equation}
\label{Ren4}
D_{1,1}=-8\left[\frac{1}{64\pi^2}\frac{1}{\widetilde \epsilon}\right],
D_{2,1}=3\left[\frac{1}{64\pi^2}\frac{1}{\widetilde \epsilon}\right],
D_{3,1}=\frac{1}{64\pi^2}\frac{1}{\widetilde \epsilon}.
\end{equation}

For all steps we always have
\begin{equation}
\label{Ren5}
A_n=A_1, \overline{A}_n =\overline{A}_1, E_n=E_1, B_n=B_1, \overline{B}_n =\overline{B}_1, F_n=F_1, C_n=C_1, D_{1,2,3,n}= D_{1,2,3,1}
\end{equation}
and for $n>1$
\begin{equation}
\label{Ren6}
\overline{\overline A}_n = -1 + A_1, \overline{\overline B}_n = -1 + B_1 , \overline{\overline C}_n = -1 - C_1
\end{equation}

\subsection{Comments related with the presence of the fermions in the game}

We must remark that, on eqs. (\ref{pio4}), (\ref{sig5}) and (\ref{p6}) of appendix \ref{apend2}, the following terms: $\frac{8g^2}{(4\pi)^2}m_\psi ^2 \frac{1}{\widetilde \epsilon}$, $- \frac{8g^2}{(4\pi)^2}3m_\psi ^2 \frac{1}{\widetilde \epsilon}$, $ -\frac{g^2}{(4\pi)^2}\left[ \frac{ m_\sigma ^2}{2k_0}\right] \frac{1}{\widetilde \epsilon}$ and $ -\frac{g^2}{(4\pi)^2} \left[3 \frac{ m_\pi ^2}{2k_0}\right] \frac{1}{\widetilde \epsilon}$ should be neglected. As stated by the authors of \cite{Karsch} and remarked by the authors of \cite{Chiku}, these terms will be canceled by contributions from higher order loops. Since we are concerned only about the one-loop approximation, we do not have to worry about them. Nevertheless, in the $O(4)$ linear sigma model, i.e., when $g=0$, none of the above terms will be present, and our model will be order by order renormalizable in the regions of validity of the MSCR. This occurs because our tree-level resummed masses are related by a symmetry relation that always guarantees the cancellation of the UV divergences.  

\section{Numerical Analysis}
\label{nun}

In this section, we present numerical solutions of the gap equations for the tree-level meson and fermion masses and the condensate derived in section \ref{loop3} including all diagrams which belong to the one loop order.

As an approximation, only for the sake of obtaining continuous curves, in the numerical evaluation we considered $\Delta\Pi=0$ also in the intermediate 
temperature region. Rigorously speaking, the curves should only be trusted in 
the low and high temperature regions. Figure \ref{six} shows the tree-level resummed meson masses, eqs. (\ref{p10}) and (\ref{p11}), as functions of the temperature. We show in Figure \ref{seven} the tree-level fermion resummed mass, eq. (\ref{p12}), as a function of temperature. The tree-level masses behavior 
exhibit the fact that the MSCR has solved the problem of tachyonic masses. In Figure \ref{eight} the chiral condensate $\nu$, eq.(\ref{p13}), 
as a function of temperature is shown whereas in Figure \ref{nine} the 
condensate is ploted in the case $M_\pi=0$.

Since at low temperatures the condensate dominates, the mesons masses suffers its influence in this region. The sigma mass decreases and they approach each other to become degenerate in a temperature of about $300 MeV$. This confirms the results we found in a phenomenological approach to the linear sigma model\cite{Caldas}.

The condensate is a slowly decreasing function of the temperature, which is a signature of the order parameter when the symmetry breaking term is present. The qualitative behavior of the results shown in Figs. \ref{six} and \ref{eight} can be compared with the ones obtained by Chiku and Hatsuda \cite{Chiku} since OPT also sums three-point vertex diagrams, as our method does. Some differences may be attributed to the incorporation of the fermions, as performed in our method. In the absence of the chiral symmetry breaking term, i.e., when $c=0$, the non-vanishing solutions of the extremum condition, eq. (\ref{derpot1}), are obtained numerically by equation (\ref{p13}) with $M^2_{\pi}=0$, $M^2_{\sigma}=2\lambda\nu^2$ and $M_\psi= M_\psi(M^2_{\pi}=0, M^2_{\sigma}=2\lambda\nu^2)$. The solution is depicted in Fig. \ref{nine} and gives an indication of first order phase transition. This result agrees with the predictions of first order phase transition found in previous analysis by Roh and Matsui\cite{Roh1}, Petropoulos\cite{Petropoulos}, Chiku and Hatsuda\cite{Chiku}, Randrup\cite{Randrup} and Bilic\cite{Bilic}. Of course we 
have to bear in mind that our result is at one loop order in the perturbative expansion. It may well be that near the critical temperature higher order corrections become crucial and may change the order of the phase transition.

The tree fermion mass, Fig. \ref{seven}, does not become zero when chiral symmetry is restored and $\nu\to0$ since we considered contributions from the mesons, given by eq. (\ref{self7}). On the contrary, when the temperature is $\geq 200 MeV$ these contributions dominate the variable $\nu$ and the fermion mass increases with temperature. The behavior of the fermion mass is in agreement with the results found by Panda in \cite{Panda} for the quark meson coupling model.

\section{Concluding Remarks}
\label{conc} 
In this paper, we presented a modified self-consistent resummation (MSCR) at finite temperature. Results for the chiral fermion meson model and the 
massless $\lambda \phi^4$ model in the weak coupling limit were obtained and analyzed. We have shown that our procedure properly resumes higher order 
terms which cures the problem of the breakdown of the perturbative expansion.

We have also shown that the MSCR, when applied to the study of the chiral fermion meson model, has the essential features which leads to the satisfaction 
of Goldstone's theorem and renormalization of the UV divergences, in the 
low and high temperature regions. We have explicitly shown that the scheme breaks down around $T_c$ i.e., in the region of intermediate temperatures. 
The application of the MSCR in these three physically different regions 
(low, intermediate and high temperatures) revealed a source of mistakes 
usually found in the literature, that is to treat all ranges of temperatures 
in the same way. It is naive to expect that the same approximations which is valid, e.g., for high temperatures would be enough in the intermediate temperature region, since quantum fluctuations are known to play a major role there.

This division was essential to identify the regions where higher order terms and resummation are crucial. It is valid to remember that even when higher order loops are taken into account, the resummation is still necessary since the tree-level masses will become tachyonic even below the critical temperature (in theories with spontaneous symmetry breaking) and break the perturbative expansion. This breakdown of the perturbative expansion can also happen in massless field theories, like QCD, due to the appearance of infrared divergences. As we discussed, the breakdown of perturbative expansion in finite temperature field theory requires resummation techniques as the MSCR to recover the reliability of perturbative expansion. 

In each region renormalization and satisfaction of Goldstone's theorem were discussed in detail. In our study, we have also addressed a usually avoided point: the inclusion of the fermions. Finally, we have re-examined the chiral phase transition in static equilibrium in terms of the linear sigma model with our MSCR.

The gap equations for the tree-level masses, are constructed by our method and in the effective Lagrangian they are renormalized. For the particular case of 
intermediate temperatures region, the gap equations would be renormalized in 
the (reorganized) effective Lagrangian only if $\Delta\Pi=\Pi_\sigma-\Pi_\pi=0$. In most of the approximations found in the literature, the gap equations are reached by some technique or via some ad-hoc procedure but the Lagrangian is yet the original one. This makes the renormalization process non-trivial, unless a finite 
cut-off is used and the theory is treated as an effective model \cite{Camelia3,Lenaghan}. As pointed out by Chiku and Hatsuda \cite{Chiku}, 
the resummation must be done also in the counterterms, which is essential to show the renormalization.

At this point, it is extremely worth emphasizing that, although one has the freedom of adding and subtracting mass parameters to the Lagrangian, in this case they can not be completely arbitrary. If the mass parameters introduced were different for the pion and sigma fields (i.e., $\frac{1}{2}M_1 \sigma^2+\frac{1}{2}M_2 \vec\pi^2$ and, of course, the same quantities subtracted, with $M_1 \ne M_2$), neither the $O(4)$ linear $\sigma$ model is renormalizable in any given order nor Goldstone´s theorem is satisfied. This will happen even if the mass parameters are determined by some physical condition as FAC or principle of minimal sensitivity (PMS). So, the most important fact behind the fulfillment of Goldstone's theorem and renormalizability of theories with SSB is the chiral symmetry that must 
dictate which mass parameter should be introduced to the Lagrangian.

\section*{Acknowledgements}
One of the authors (H. C. G. Caldas) thanks the hospitality given by the 
Nuclear Theory group during his visit at the University of Minnesota were 
part of this work was done. He is gratefully indebted to Professors J. I. Kapusta 
and P. J. Ellis for various helpful advices and enlightening discussions. 
He is also grateful to Dr. M. Hott for valuables conversations about this problem, Professor A. Das for comments concerning the effective potential 
and finally Drs. J. Lenaghan and S. Chiku for useful e-conversations. The 
authors would like to thank B. Hiller and A. Blin for useful comments on 
the subject of the paper. H. C. G. Caldas thanks the generous support 
provided by the Faculty of UFMG and FUNREI.

\appendix
\section{Renormalization of the effective potential}
\label{apend1}

As mentioned earlier the vacuum contribution to $\Omega_{1} (T, \nu)$ is divergent and requires renormalization. In this subsection we also use dimensional regularization in the computation of the effective potential.

\begin{equation}
\label{re1}
\Omega_{1}^0(m) \equiv \int \frac {d^3p} {(2\pi)^3} \frac{\omega}{2}
\end{equation}

\begin{equation}
\label{re2}
\frac{\partial \Omega_{1}^0 (m)} {\partial m}=m \int \frac {d^3p} {(2\pi)^3} \frac{1}{2\omega}=mL(m)
\end{equation}
where $L(m)$ is the usual zero temperature loop integral

\begin{equation}
\label{re3}
L(m)\equiv \int \frac {d^4p} {(2\pi)^4} \frac{1}{p^2+m^2}=\int \frac {d^4p} {(2\pi)^4} \frac{1}{p_4^2+\bp^2+m^2},
\end{equation}
with $d^4p=dp_4d^3p$ being the four Euclidean momentum.

The divergent integral $L(m)$ can be evaluated in the standard manner

\begin{equation}
\label{re4}
\frac{ m ^2}{(4\pi)^2} \left[- \frac{1}{\widetilde \epsilon} -1 + \ln \left(\frac{m ^2}{\mu^2}\right) \right].
\end{equation}

The quantity $\Omega_{1}^0(m)$ is then obtained with the integration 
of $m L\left(m\right)$

\begin{equation}
\label{re8}
\Omega_{1}^0(m)= \frac{m^4}{64\pi^2}(\ln \frac{m^2}{\mu^2}-\frac{3}{2}- \frac{1}{\widetilde \epsilon})
\end{equation}

With this expression we can find the zero temperature effective potential,

\begin{equation}
\label{re9}
\Omega_{1}^0(\nu)=-\frac{m_{\sigma}^4}{64\pi^2}\frac{1}{\widetilde \epsilon}
+\frac{m_{\sigma}^4}{64\pi^2}(\ln \frac{m_{\sigma}^2}{\mu^2}-\frac{3}{2})+ 3(m_{\sigma} \leftrightarrow m_{\pi})-8(m_{\sigma} \leftrightarrow m_{\psi})
\end{equation} 

The renormalization of the thermodynamical potential at the end amounts to the determination of the parameters $D_{1,2,3}$,

\begin{equation}
\label{re10}
D_1=-8\left[\frac{1}{64\pi^2}\frac{1}{\widetilde \epsilon}\right],
\end{equation}

\begin{equation}
\label{re11}
D_2=3\left[\frac{1}{64\pi^2}\frac{1}{\widetilde \epsilon}\right],
\end{equation}

\begin{equation}
\label{re12}
D_3=\frac{1}{64\pi^2}\frac{1}{\widetilde \epsilon}.
\end{equation}

\section{One-loop self-energy at finite temperature }
\label{apend2}

At zero momentum the expressions for the self-energies are given by

\begin{equation}
\label{pio1}
\Pi_{\pi 1}(k_0, \left|\bk\right|=0)=\Pi_{\pi 1} ^0 + \Pi_{\pi 1} ^{\beta} = \frac{5\lambda}{(4\pi)^2} m_\pi ^2 \left[- \frac{1}{\widetilde \epsilon} -1 + \ln \left(\frac{m_\pi ^2}{\mu^2}\right) \right]+
\frac{5\lambda}{2}\int_0^{\infty }\frac{dp p^2}{\pi^2} \frac {n_{\pi}}{\omega_{\pi}}
\end{equation}

\begin{equation}
\label{pio2}
\Pi_{\pi 2}(k_0, \left|\bk\right|=0)=\Pi_{\pi 2} ^0 + \Pi_{\pi 2} ^{\beta} = \frac{\lambda}{(4\pi)^2}m_\sigma ^2 \left[- \frac{1}{\widetilde \epsilon} -1 
+ \ln \left(\frac{m_\sigma ^2}{\mu^2}\right) \right] +\frac{\lambda}{2}\int_0^{\infty }\frac{dp p^2}{\pi^2} \frac {n_{\sigma}}{\omega_{\sigma}}
\end{equation}

\begin{eqnarray}
\label{pio3}
\Pi_{\pi 3}(k_0, \left|\bk\right|=0)=\Pi_{\pi 3} ^0 + \Pi_{\pi 3} ^{\beta} = \\
\nonumber
\frac{4\lambda^2\nu^2}{(4\pi)^2} \left[- \frac{1}{\widetilde \epsilon} -1 + 
\ln \left(\frac{m_\pi ^2}{\mu^2}\right)+\frac{k_0 ^2+m_\sigma^2-m_\pi^2}{2k_0 ^2}\ln \left(\frac{m_\pi ^2}{m_\sigma^2}\right) +\frac{\sqrt{\Delta_3}}{k_0 ^2}(\Delta_4+\Delta_5) \right]+\\
\nonumber
-2 \lambda^2 \nu^2 \int_0^{\infty}\frac{dp p^2}{\pi^2}\left[\frac {n_{\pi}}{\omega_{\pi}} \frac{-k_0^2+m_{\sigma}^2- m_{\pi}^2}{(-k_0^2+ m_{\sigma}^2- m_{\pi}^2)^2 -2 k_0^2(\omega_{\pi}^2+ \omega_{\sigma}^2)} + \frac {n_{\sigma}}{\omega_{\sigma}} \frac{-k_0^2+ m_{\pi}^2-m_{\sigma}^2}{(-k_0^2 +m_{\pi}^2-m_{\sigma}^2)^2 -2 k_0^2(\omega_{\pi}^2+ \omega_{\sigma}^2)}\right]
\nonumber 
\end{eqnarray}

\begin{eqnarray}
\label{pio4}
\Pi_{\pi 4}(k_0, \left|\bk\right|=0)=\Pi_{\pi 4} ^0 + \Pi_{\pi 4} ^{\beta} = 
\frac{8g^2}{(4\pi)^2}\left[m_\psi ^2 - \frac{k_0 ^2}{2} \right] \frac{1}{\widetilde \epsilon} +\\
\nonumber
\frac{8g^2}{(4\pi)^2}\left[-\left(m_\psi ^2 - 
\frac{k_0 ^2}{2} \right)\ln \left( \frac{m_\psi ^2}{\mu ^2}\right)+\Delta_2\sqrt{\Delta_1}-K_0 ^2 \right]+\\
4g^2\int_0^{\infty}\frac{dp p^2}{\pi^2}\frac {n_{\psi}}{\omega_{\psi}}\left[1+ \frac{k_0^2}{ 4\omega_{\psi^2}- k_0^2}\right]
\nonumber
\end{eqnarray}

\begin{equation}
\label{sig1}
\Pi_{\sigma 1}(k_0, \left|\bk\right|=0)=\Pi_{\sigma 1} ^0 + \Pi_{\sigma 1} ^{\beta} = \frac{3\lambda}{(4\pi)^2}m_\sigma ^2 \left[- \frac{1}{\widetilde \epsilon} -1 + \ln \left(\frac{m_\sigma ^2}{\mu^2}\right)\right]+\frac{3\lambda}{2}\int_0^{\infty }\frac{dp p^2}{\pi^2} \frac {n_{\sigma}}{\omega_{\sigma}}
\end{equation}

\begin{equation}
\label{sig2}
\Pi_{\sigma 2}(k_0, \left|\bk\right|=0)=\Pi_{\sigma 2} ^0 + \Pi_{\sigma 2} ^{\beta} = \frac{3\lambda}{(4\pi)^2}m_\pi ^2 \left[- \frac{1}{\widetilde \epsilon} -1 + \ln \left(\frac{m_\pi ^2}{\mu^2}\right)\right]+\frac{3\lambda}{2}\int_0^{\infty }\frac{dp p^2}{\pi^2} \frac {n_{\pi}}{\omega_{\pi}}
\end{equation}

\begin{eqnarray} \label{sig3}
\Pi_{\sigma 3}(k_0, \left|\bk\right|=0)=\Pi_{\sigma 3} ^0 + \Pi_{\sigma 3} ^{\beta} = 18\frac{\lambda^2\nu^2}{(4\pi)^2}\left[- \frac{1}{\widetilde \epsilon} -2 + \ln \left(\frac{m_\sigma ^2}{\mu^2}\right)+2f_1(k_0)arctan\left(\frac{1}{f_1(k_0)}\right)\right]+\\
-18\lambda^2 \nu^2 \int_0^{\infty}\frac{dp p^2}{\pi^2}\frac {n_{\sigma}}{\omega_{\sigma}} \frac{1}{4\omega_{\sigma}^2- k_0^2}
\nonumber
\end{eqnarray}

\begin{eqnarray} \label{sig4}
\Pi_{\sigma 4}(k_0, \left|\bk\right|=0)=\Pi_{\sigma 4} ^0 + \Pi_{\sigma 4} ^{\beta} = 6\frac{\lambda^2\nu^2}{(4\pi)^2}\left[- \frac{1}{\widetilde \epsilon} -2 + \ln \left(\frac{m_\pi ^2}{\mu^2}\right)+2f_2(k_0)arctan\left(\frac{1}{f_2(k_0)}\right)\right]+\\
-6 \lambda^2 \nu^2 \int_0^{\infty}\frac{dp p^2}{\pi^2}\frac {n_{\pi}}{\omega_{\pi}} \frac{1}{4\omega_{\pi}^2-k_0^2}
\nonumber
\end{eqnarray}

\begin{eqnarray} \label{sig5}
\Pi_{\sigma 5}(k_0, \left|\bk\right|=0)=\Pi_{\sigma 5} ^0 + \Pi_{\sigma 5} ^{\beta} = - \frac{8g^2}{(4\pi)^2}\left[3m_\psi ^2 - \frac{k_0 ^2}{2} \right] \frac{1}{\widetilde \epsilon} +\\
\nonumber
 \frac{8g^2}{(4\pi)^2}\left[\frac{1}{2}(6m_\psi ^2-k_0 ^2)\ln \left(\frac{m_\psi ^2}{\mu^2}\right)+ (4m_\psi ^2-k_0 ^2)\left(\frac{\Delta_2 \sqrt{\Delta_1}}{k_0 ^2}-1\right)\right] + \\
4g^2\int_0^{\infty}\frac{dp p^2}{\pi^2}\frac {n_{\psi}}{\omega_{\psi}}[1+ \frac{4m_{\psi^2}-k_0^2}{ k_0^2-4\omega_{\psi^2}}],
\nonumber
\end{eqnarray}
where $f_1 (k_0)=\sqrt{\frac{4m_\sigma ^2}{k_0 ^2}-1}$, $f_2 (k_0)=\sqrt{\frac{4m_\pi ^2}{k_0 ^2}-1}$, $\Delta_1=k_0 ^2(k_0 ^2-4m_\psi ^2)$, $\Delta_2=arctan\left(\frac{1}{\sqrt{1-\frac{4m_\psi ^2}{k_0 ^2}}}\right)$, $\Delta_3=k_0 ^4-2k_0 ^2(m_\pi ^2+m_\sigma ^2)+ (m_\pi ^2-m_\sigma ^2)^2$, $\Delta_4=arctanh\left(\frac{k_0 ^2+(m_\pi ^2+m_\sigma ^2)}{\sqrt{\Delta_3}} \right)$ and $\Delta_5=arctanh\left(\frac{k_0 ^2-(m_\pi ^2+m_\sigma ^2)}{\sqrt{\Delta_3}}\right)$. 

\begin{eqnarray} \label{p6}
\Sigma(k_0, \left|\bk\right|=0)= (\Sigma_0+\Sigma_s)_\sigma +3(\Sigma_0+\Sigma_s)_\pi=(\Sigma_0 ^0+\Sigma_0 ^\beta+\Sigma_s ^0+\Sigma_s ^\beta)_\sigma +3(\Sigma_0 ^0+\Sigma_0 ^\beta+\Sigma_s ^0+\Sigma_s ^\beta)_\pi=\\
\nonumber
\frac{g^2}{(4\pi)^2}\left[m_\psi+ \frac{1}{2k_0}(k_0 ^2+m_\psi ^2-m_\sigma ^2)\right] \frac{1}{\widetilde \epsilon}+\\
\nonumber
-\frac{g^2}{(4\pi)^2}m_\psi \left[\ln \left(\frac{m_\psi ^2}{\mu^ 2} \right)+Z\right] - 
\frac{1}{(4\pi)^2}\frac{g^2}{2k_0}(k_0 ^2+m_\psi ^2-m_\sigma ^2) \left[\ln \left(\frac{m_\psi ^2}{\mu^ 2} \right)+Z\right]+\\
\nonumber
\frac{g^2}{2}\int_0^{\infty }\frac{dp p^2}{\pi^2} \frac {n_{\sigma}}{\omega_{\sigma}} \frac{[k_0(-k_0^2+\omega_\sigma^2+\omega_\psi^2)+m_\psi(-k_0^2-\omega_\sigma^2+\omega_\psi^2)]}{[ k_0^2-( \omega_\psi -\omega_\sigma)^2] [ k_0^2-( \omega_\psi +\omega_\sigma)^2]} 
\nonumber +\\
\frac{g^2}{2}\int_0^{\infty }\frac{dp p^2}{\pi^2} \frac {n_{\psi}}{\omega_{\psi}} \frac{[2k_0\omega_\psi^2+m_\psi(k_0^2-\omega_\sigma^2+\omega_\psi^2)]}{[ k_0^2-( \omega_\psi -\omega_\sigma)^2] 
[ k_0^2-( \omega_\psi +\omega_\sigma)^2]} +3(m_\sigma \leftrightarrow m_\pi)
\nonumber
\end{eqnarray}
where $\Sigma_{s}(T,k_0, \left|\bk\right|=0)$ is the scalar contribution, proportional to the unit matrix and $\Sigma_{0}(T,k_0, \left|\bk\right|=0)$ 
is the contribution proportional to the matrix $\gamma^0$, and

\begin{eqnarray}
\label{fer}
\left( \Sigma_s ^\beta \right)_\sigma = \frac{g^2m_\psi}{2\pi^2} \int_0^{\infty }dp p^2 \left[\frac {n_{\sigma}}{\omega_{\sigma}}\frac{m_\psi ^2-m_\sigma ^2 -k_0^2}{[ k_0^2-( \omega_\psi -\omega_\sigma)^2] [ k_0^2-( \omega_\psi +\omega_\sigma)^2]}  + \right. \\
\left. \frac{n_{\psi}}{\omega_{\psi}}\frac{m_\psi ^2-m_\sigma ^2 +k_0^2}{[ k_0^2-( \omega_\psi -\omega_\sigma)^2] [ k_0^2-( \omega_\psi +\omega_\sigma)^2]} \right]
\nonumber
\end{eqnarray}

\begin{eqnarray}
\label{fer0}
\left( \Sigma_0 ^\beta \right)_\sigma = \frac{g^2 k_0}{2\pi^2} \int_0^{\infty }dp p^2 \left[\frac {n_{\sigma}}{\omega_{\sigma}}\frac{\omega_\psi ^2+\omega_\sigma ^2 - k_0^2}{[ k_0^2-( \omega_\psi -\omega_\sigma)^2] [ k_0^2-( \omega_\psi +\omega_\sigma)^2]}  + \right. \\
\left. \frac{n_{\psi}}{\omega_{\psi}}\frac{2 \omega_\psi ^2}{[ k_0^2-( \omega_\psi -\omega_\sigma)^2] [ k_0^2-( \omega_\psi +\omega_\sigma)^2]} \right]
\nonumber
\end{eqnarray}

\begin{equation}
\label{fer1}
\left( \Sigma_s ^0 \right)_\sigma = -\frac{g^2}{(4\pi)^2}m_\psi \left[\ln \left(\frac{m_\psi ^2}{\mu^ 2} \right)+Z\right]
\end{equation}

\begin{equation}
\label{fer2}
\left( \Sigma_0 ^{0} \right)_\sigma = -\frac{1}{(4\pi)^2}\frac{g^2}{2k_0}(k_0 ^2+m_\psi ^2-m_\sigma ^2) \left[\ln \left(\frac{m_\psi ^2}{\mu^ 2} \right)+Z\right],
\end{equation}
with $Z$ defined as

\begin{equation}
\label{fer3}
Z \equiv \frac{1}{(4\pi)^2}\left[\frac{k_0 ^2+m_\sigma^2-m_\psi^2}{2k_0 ^2}\ln \left(\frac{m_\psi ^2}{m_\sigma^2}\right) +\frac{\sqrt{\Delta_6}}{k_0 ^2}(\Delta_7+\Delta_8) \right],
\end{equation}
with $\Delta_6=k_0 ^4-2k_0 ^2(m_\psi ^2+m_{\sigma;\pi }^2)+ (m_\psi ^2-m_{\sigma ;\pi}^2)^2$, $\Delta_7=arctanh\left(\frac{k_0 ^2+(m_\psi ^2+m_{\sigma ;\pi}^2)}{\sqrt{\Delta_6}} \right)$ and $\Delta_8=arctanh\left(\frac{k_0 ^2-(m_\psi ^2+m_{\sigma ;\pi}^2)}{\sqrt{\Delta_6}}\right)$.

\newpage


\newpage

\begin{figure}[h]
\epsfxsize= 15cm
\vspace{0.8cm}
\centerline{\epsffile{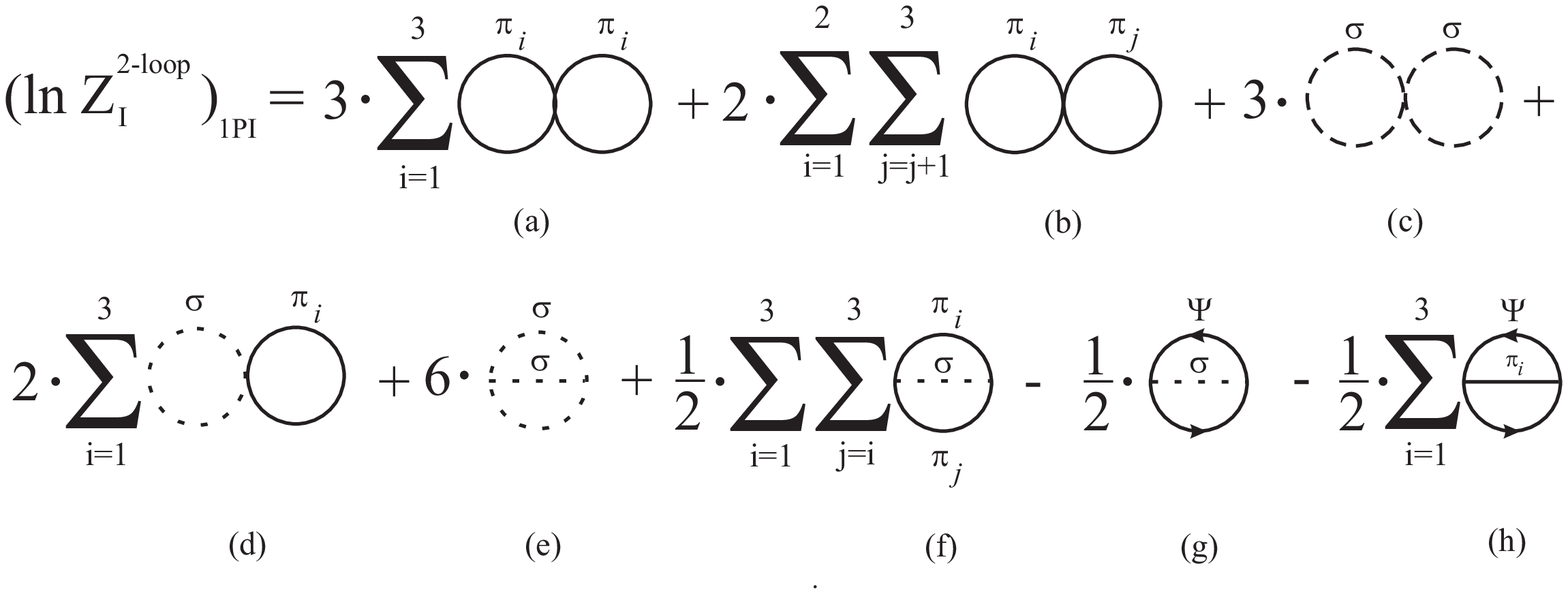} }
\vspace{0.8cm}
\protect\caption[]{Figure 1: The logarithm of the interaction partition function 1PI. Solid line is the tree-level pion meson propagator, the dashed line is the tree-level sigma meson propagator and the solid line with a arrow is the tree-level fermion propagator.}
\label{one}
\end{figure}

\newpage

\begin{figure}[h]
\vspace{0.8cm}
\centerline{\epsffile{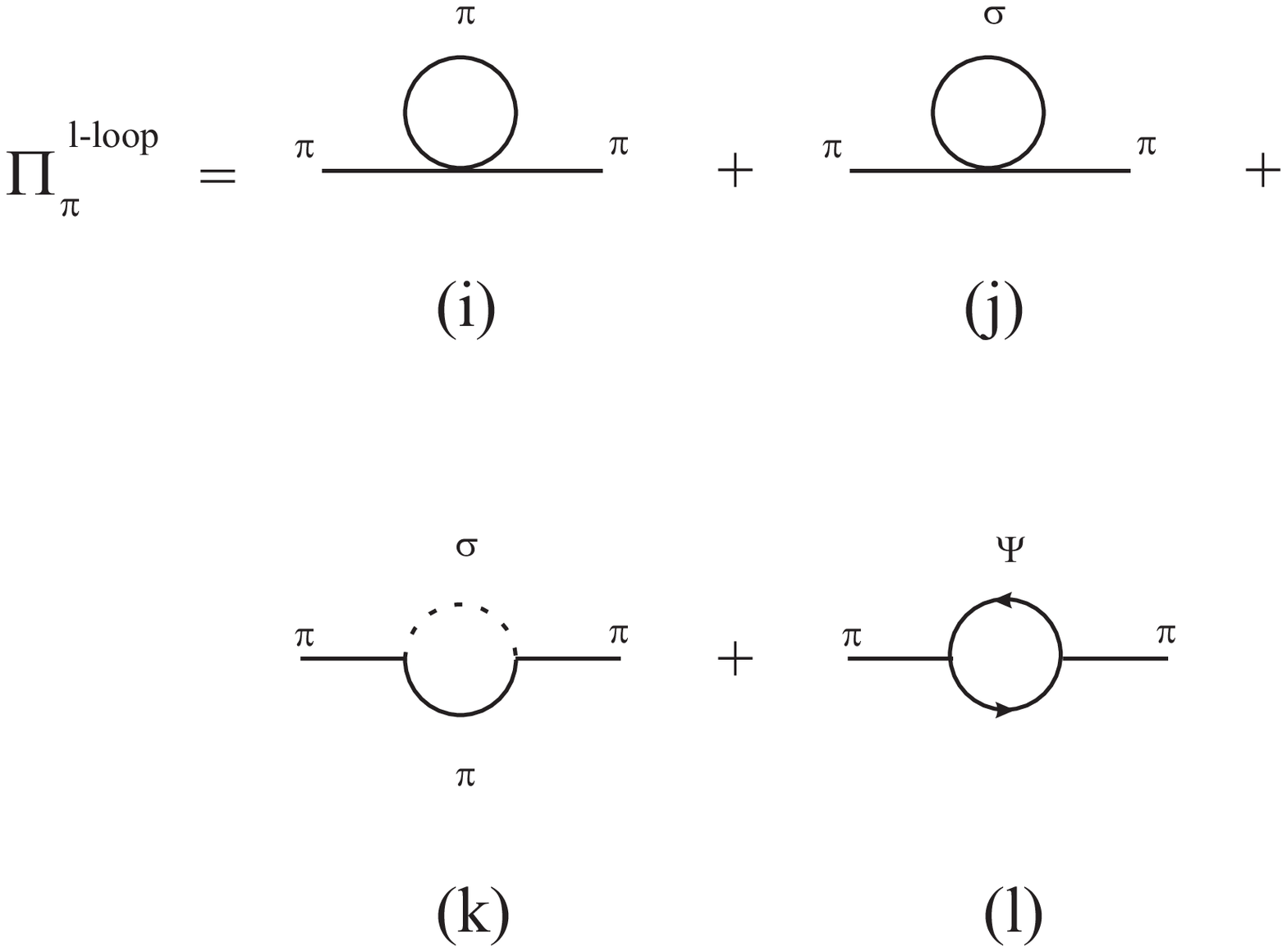} }
\vspace{0.8cm}
\protect\caption[]{Figure 2: One-loop self-energy for the pions.}
\label{two}
\end{figure}

\newpage

\begin{figure}[h]
\vspace{0.8cm}
\centerline{\epsffile{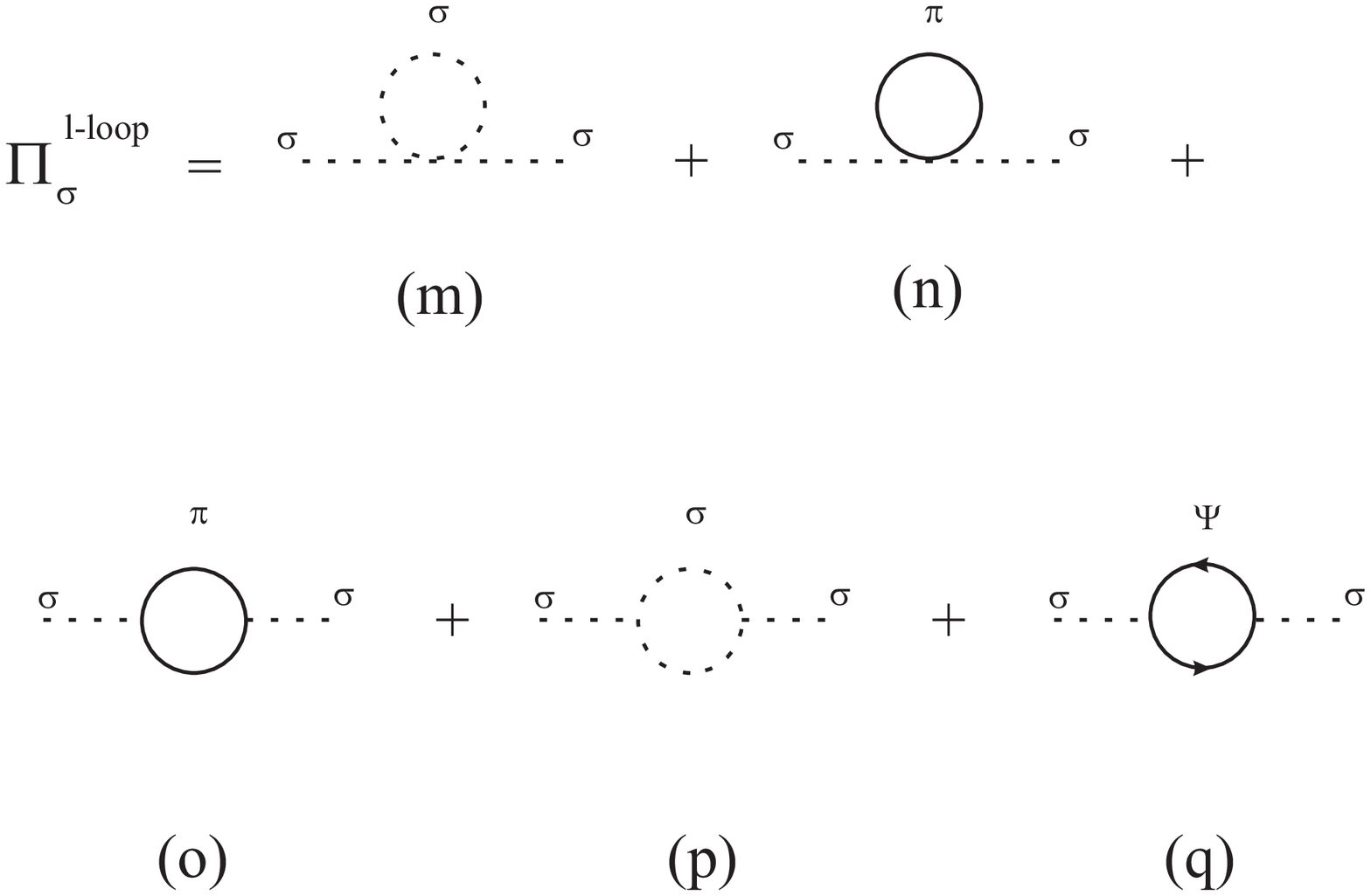} }
\vspace{0.8cm}
\protect\caption[]{Figure 3: One-loop self-energy for the meson sigma.}
\label{three}
\end{figure}

\newpage

\begin{figure}[h]
\vspace{0.8cm}
\centerline{\epsffile{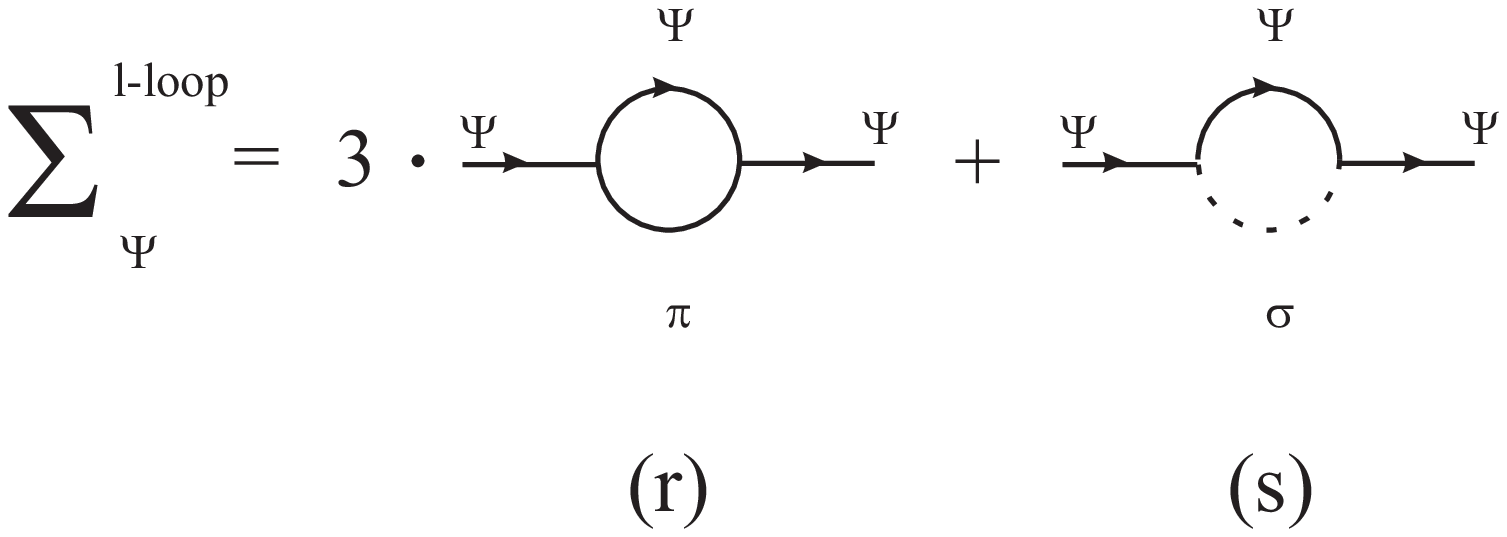} }
\vspace{0.8cm}
\protect\caption[]{Figure 4: One-loop self-energy for the fermions. Note that the lowest corrections for the fermions are of order $g^2$.}
\label{four}
\end{figure}

\newpage

\begin{figure}[h]
\vspace{0.8cm}
\centerline{\epsffile{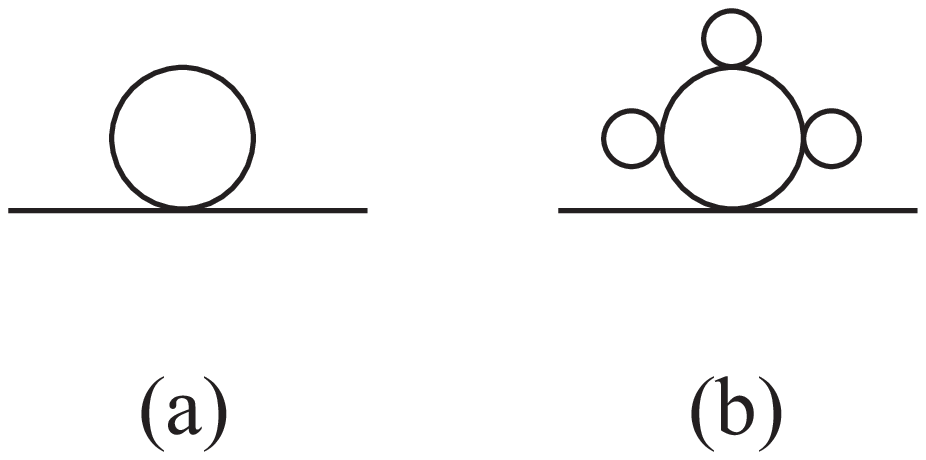} }
\vspace{0.8cm}
\protect\caption[]{Figure 5: (a) The 1PI one-loop self-energy diagram of the $\lambda \phi^4$ model and (b) a ``daisy'' type diagram with three attached bubbles which contributes to the self-energy.}
\label{five}
\end{figure}

\newpage

\begin{figure}[h]
\epsfxsize= 15cm 
\vspace{0.8cm}
\centerline{\epsffile{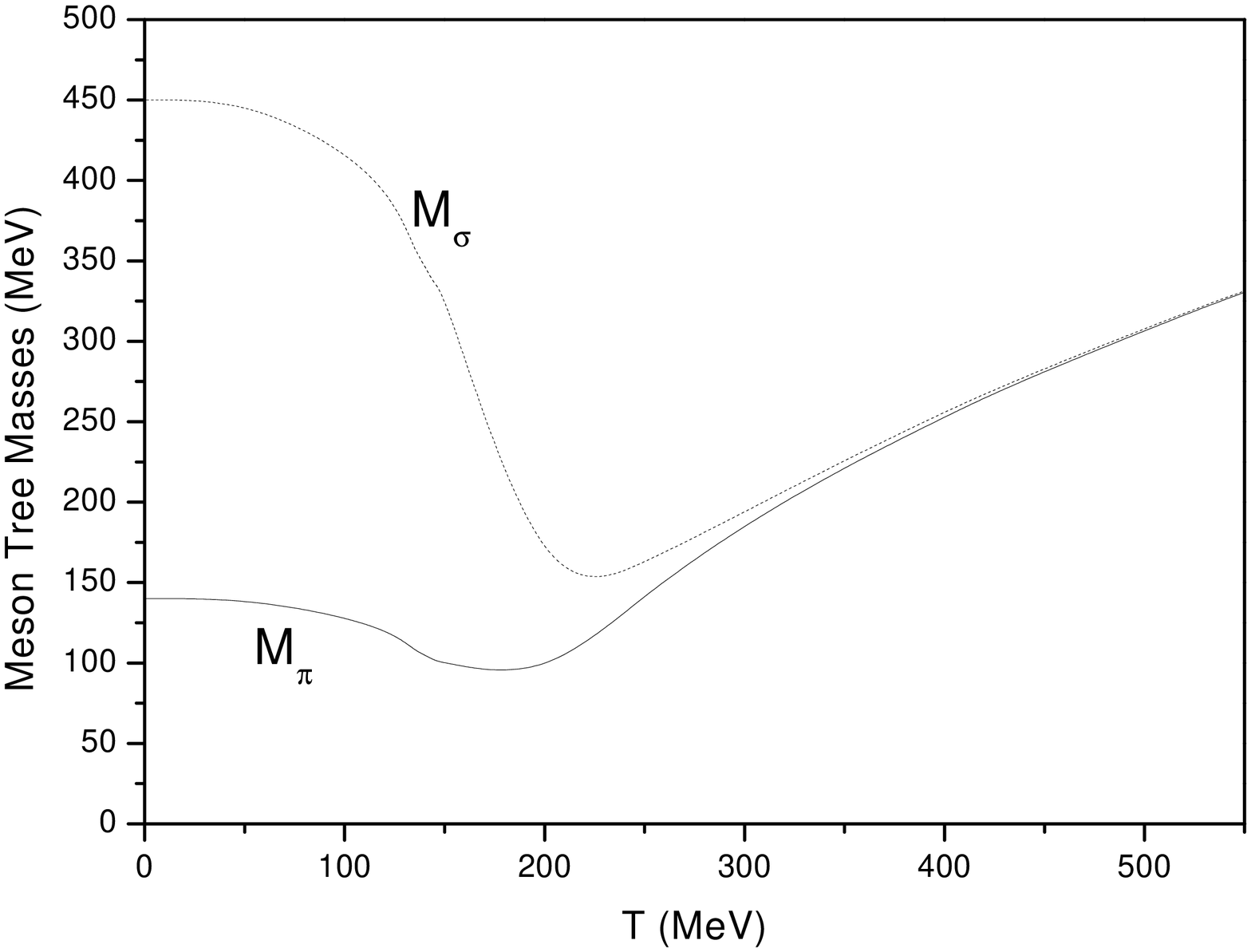} }
\vspace{0.8cm}
\protect\caption[]{Figure 6: Tree-level resumed meson masses.}
\label{six}
\end{figure}

\newpage

\begin{figure}[h]
\epsfxsize= 15cm 
\vspace{0.8cm}
\centerline{\epsffile{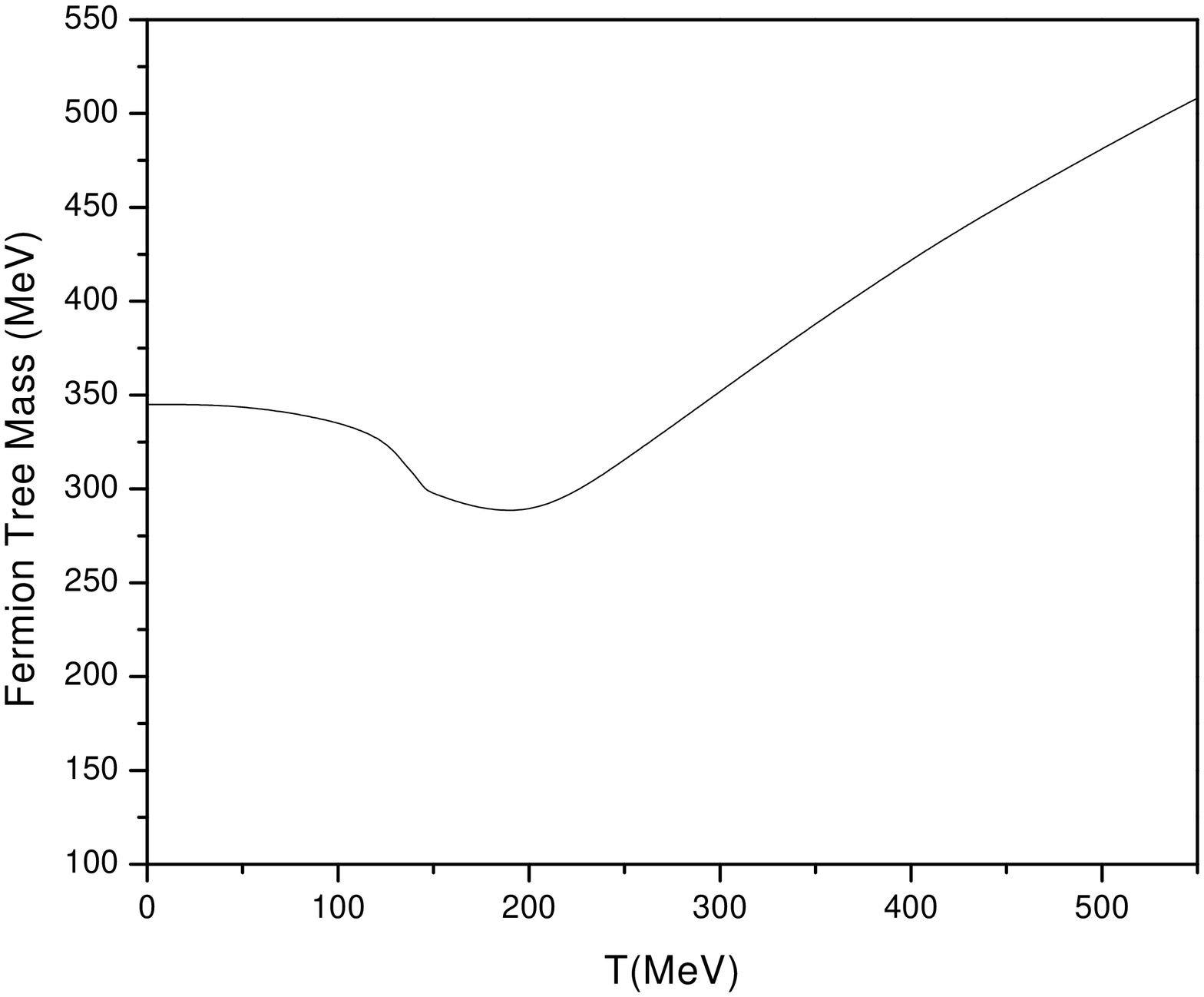} }
\vspace{0.8cm}
\protect\caption[]{Figure 7: Tree-level resumed fermion mass.}
\label{seven}
\end{figure}

\newpage

\begin{figure}[h]
\epsfxsize= 15cm 
\vspace{0.8cm}
\centerline{\epsffile{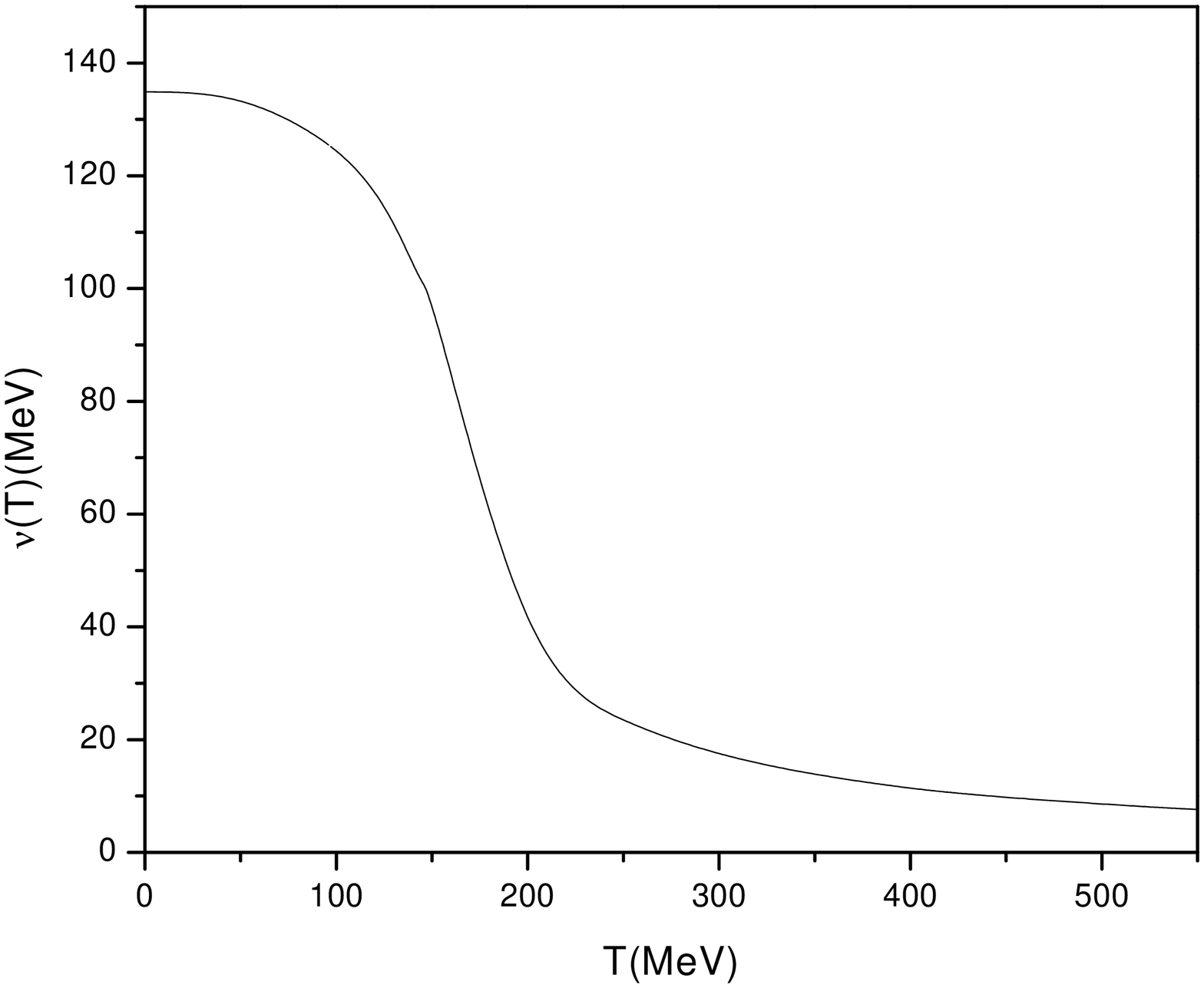} }
\vspace{0.8cm}
\protect\caption[]{Figure 8: Condensate $\nu$ as a function of the temperature.}
\label{eight}
\end{figure}

\newpage

\begin{figure}[h]
\epsfxsize= 15cm 
\vspace{0.8cm}
\centerline{\epsffile{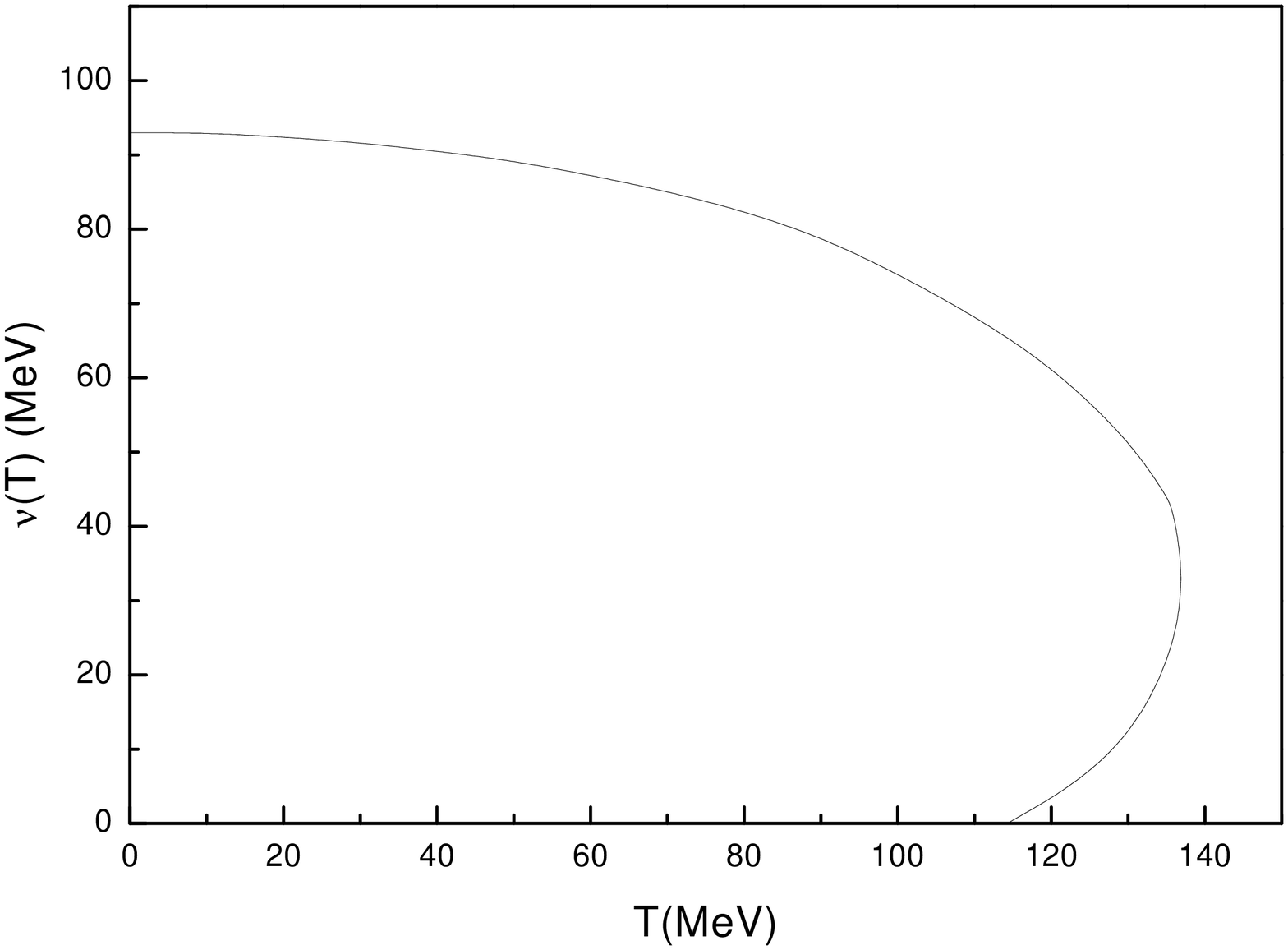} }
\vspace{0.8cm}
\protect\caption[]{Figure 9: Condensate $\nu$ as a function of the temperature in the chiral limit ($M_\pi=0$).}
\label{nine}
\end{figure}


\begin{references} 
\let\ul=\underbar
\def\AP#1{{Ann. Phys. }{\bf #1}}
\def\PRP#1{{Phys. Rep. }{\bf #1}}
\def\APP#1{{Act. Phys. Pol. }{\bf #1}}
\def\PTP#1{{Prog. Theor. Phys. }{\bf #1}}
\def\PR#1{{Phys. Rev. }{\bf #1}}
\def\PRD#1{{Phys. Rev. }{\bf D #1}}
\def\PRC#1{{Phys. Rev. }{\bf C #1}}
\def\PRL#1{{Phys. Rev. Lett. }{\bf #1}}
\def\PL#1{{Phys. Lett. }{\bf #1}}
\def\RMP#1{{Rev. Mod. Phys.}{\bf #1}}
\def\NP#1{{Nucl. Phys. }{\bf #1}}
\def\ZP#1{{Z. Phys. }{\bf #1}}
\def\NC#1{{Nuovo Cimento }{\bf #1}}
\def\SJNP#1{{Sov. J. Nucl. Phys. }{\bf #1}} 

\bibitem{Gell-Mann} M. Gell-Mann, and M. Levy, \NC{16}, 705 (1960).
\bibitem{Franco} D. H. T. Franco, H. C. G. Caldas, A. L. Mota and M. C. Nemes, Mod. Phys. Lett. A617, 464 (1997) and references therein.
\bibitem{Ram-Mohan}L. R. Ram Mohan, \PRD{14}, 2670 (1976).
\bibitem{Lattice}Proc. of Lattice'96, \NP{B 53}, 1 (1997).
\bibitem{Quark}Proc. of Quark Matter'97, \NP{A 638}, 1 (1998).
\bibitem{Dolan} L. Dolan and R. Jackiw, Phys. Rev. D {\bf 9}, 3320 (1974).
\bibitem{Weinberg} S. Weinberg, \PRD{\bf 9}, 3357 (1974).
\bibitem{Linde} A. Linde, Rep. Prog. Phys. \NC{42}, 389 (1979); D. J. Gross, R. D. Pisarski, and L. G. Yaffe, Rev. Mod. Phys. \NC{53}, 43 (1981).
\bibitem{Baym} G. Baym and G. Grinstein, \PRD{15}, 2897 (1977).
\bibitem{Mallik} N. Banerjee and S. Malik, \PRD{\bf 43}, 3368 (1991).
\bibitem{Parwani} Rajesh R. Parwani, \PRD{\bf 45}, 4695 (1992).
\bibitem{Chiku} S. Chiku and T. Hatsuda, \PRD{\bf 58}, 076001 (1998).
\bibitem{Kapusta} J. Kapusta, {\it Finite-Temperature Field Theory}
(Cambridge University Press, Cambridge, 1989).
\bibitem{Lee} B. Lee, {\it Chiral Dynamics} (Gordon and Breach, 1970).
\bibitem{Kirzhnits} D. A. Kirzhnits and A. D. Linde, \PL{42 B}, 471
(1972); \AP{101}, 195 (1976).
\bibitem{Camelia1} G. Amelino-Camelia and S.-Y. Pi, \PRD{47},
2356 (1993).
\bibitem{Camelia2} G. Amelino-Camelia, \PRD{49},
2740 (1994). 
\bibitem{Roh1}H.-S. Roh and Matsui T., Eur Phys J A1 205 (1998).
\bibitem{Petropoulos} N. Petropoulos, J. Phys. G25 2225 (1999).
\bibitem{Pierre} P. Ramond, {\it Field Theory: a modern primer}
(Addison Wesley Publ. Co., 1990).
\bibitem{Bernard} C. W. Bernard, \PRD {\bf 9}, 3312 (1974).
\bibitem{Bochkarev} A. Bochkarev and J. I. Kapusta, \PRD{\bf 54}, 4066 (1996).
\bibitem{Landshoff}P.V. Landshoff, hep-ph/9808362.
\bibitem{Chanowitz}M. Chanowitz, M. Furman, and I. Hinchliffe, \NP{B 159}, 225 (1979).
\bibitem{Arnold}Peter Arnold and Olivier Espinosa. \PRD{\bf 47}, 3546 (1993).
\bibitem{Cornwall}J. M. Cornwall, R. Jackiw, and E. Tomboulis,
\PRD{10}, 2428 (1974); G. Amelino-Camelia and S.-Y. Pi, \PRD{\bf 47},
2356 (1993).
\bibitem{Camelia3} G. Amelino-Camelia, \PL{B407}, 268
(1994). 
\bibitem{Lenaghan} J. T. Lenaghan and D. H. Rischke, J. Phys. G26, 431 (2000).
\bibitem{Karsch} F. Karsch, A. Patkós and P. Petreczky, \PL{B401}, 69 (1997).
\bibitem{Caldas} H. C. G. Caldas, D. H. T. Franco, A. L. Mota, F. A. Oliveira and M. C. Nemes, \NP{A 617}, 464 (1997).
\bibitem{Randrup}J. Randrup, \PRD{\bf 55}, 1188 (1997).
\bibitem{Bilic} N. Bilic and H. Nikolie, Eur. Phys. J. C6 513 (1999).
\bibitem{Panda} P. K. Panda, A. Mishra, J. M. Eisemberg and W. Greiner, \PRC{\bf 56}, 3134 (1997).

\end{references}
\end{document}